\documentclass[fleqn,usenatbib]{rasti}

\usepackage{newtxtext,newtxmath}

\usepackage[T1]{fontenc}

\DeclareRobustCommand{\VAN}[3]{#2}
\let\VANthebibliography\thebibliography
\def\thebibliography{\DeclareRobustCommand{\VAN}[3]{##3}\VANthebibliography}

\usepackage{graphicx}	
\usepackage{amsmath}	

\usepackage{listings}
\usepackage{xcolor}

\definecolor{codegreen}{rgb}{0,0.6,0}
\definecolor{codegray}{rgb}{0.5,0.5,0.5}
\definecolor{codepurple}{rgb}{0.58,0,0.82}
\definecolor{backcolour}{rgb}{0.95,0.95,0.92}

\lstdefinestyle{mystyle}{
    backgroundcolor=\color{backcolour},   
    commentstyle=\color{codegreen},
    keywordstyle=\color{magenta},
    numberstyle=\tiny\color{codegray},
    stringstyle=\color{codepurple},
    basicstyle=\ttfamily\footnotesize,
    breakatwhitespace=false,         
    breaklines=true,                 
    captionpos=b,                    
    keepspaces=true,                 
    numbers=left,                    
    numbersep=5pt,                  
    showspaces=false,                
    showstringspaces=false,
    showtabs=false,                  
    tabsize=2
}

\usepackage{url}
\usepackage{tabularx}

\newcommand{\phoptic}{\texttt{phoptic}}

\title[\phoptic{}]{\phoptic{} - a Python package for reducing astronomical images}

\author[Z. A. Irving et al.]{
Z. A. Irving,$^{1}$\thanks{E-mail: z.irving@soton.ac.uk}
N. Castro Segura,$^{2}$
D. Altamirano,$^{1}$\thanks{E-mail: d.altamirano@soton.ac.uk}
F. Vincentelli,$^{3,1}$
A. Castro,$^{4}$
R. P. Petrucci,$^{5, 6}$
\newauthor
R. Michel,$^{4}$
A. B. Hill,$^{7,1}$
J. V. Hern\'andez Santisteban$^{8}$
\\
$^{1}$School of Physics and Astronomy, University of Southampton, University Road, Southampton SO17 1BJ, UK\\
$^{2}$Department of Physics, University of Warwick, Gibbet Hill Road, Coventry CV4 7AL, UK\\
$^{3}$Fluid and Complex Systems Centre, Coventry University, CV1 5FB, UK\\
$^{4}$Instituto de Astronom\'ia, Universidad Nacional Aut\'onoma de M\'exico, Carretera Tijuana-Ensenada Km. 107, Ensenada, 22860, M\'exico\\
$^{5}$Universidad Nacional de C\'ordoba, Observatorio Astron\'omico de C\'ordoba, Laprida 854, C\'ordoba X5000BGR, Argentina\\
$^{6}$Consejo Nacional de Investigaciones Cient\'ificas y T\'echnicas (CONICET), Godoy Cruz 2290, CPC 1425FQB CABA, Argentina\\
$^{7}$ComplyAdvantage, 2nd Floor, Fetter Yard, Fetter Lane, London EC4A 1AD, United Kingdom\\
$^{8}$SUPA Physics and Astronomy, University of St Andrews, KY16 9SS, Scotland, UK
}

\date{Accepted XXX. Received YYY; in original form ZZZ}

\pubyear{\the\year{}}

\begin{document}
\label{firstpage}
\pagerange{\pageref{firstpage}--\pageref{lastpage}}
\maketitle

\begin{abstract}
Publicly-available photometry pipelines make astronomical data reduction accessible to non-experts, reduce the margin for human error, and enable reproducible reduction. In many cases, bespoke reduction software is written on a per-instrument basis; this results in rigid pipelines that cannot be straightforwardly applied to data from other instruments. To alleviate this problem, we present \phoptic{}, an open source photometry pipeline written in Python. \phoptic{} began as a dedicated pipeline for the the OPtical TIming CAMera (OPTICAM), a triple-camera system mounted on the 2.1~m telescope at the Observatorio Astron\'omico Nacional in San Pedro M\'artir, M\'exico. However, \phoptic{} now serves as a generic photometry pipeline with a simple interface to reduce data from other instruments. At its core, \phoptic{} leverages the \texttt{astropy} Python package, and affiliated packages thereof, to provide a flexible, modern, and interoperable reduction pipeline. In particular, \phoptic{} uses \texttt{photutils} for background estimation, source detection, and performing aperture photometry. Additionally, \phoptic{} implements optimal photometry, improving the signal-to-noise ratio over aperture photometry by up to $\sim 10$ per cent. We describe \phoptic{}'s functionality, discuss its default behaviour, and demonstrate its flexible interface by reducing data from the HiPERCAM, MEXMAN, OPTICAM, and ULTRACAM instruments. We also review the performance of \phoptic{}, and show that it is highly scalable on multi-core CPUs.
\end{abstract}




\section{Introduction}\label{sec: intro}

Optical light curves (i.e., measurements of an astrophysical source's optical brightness over time) are one of the most fundamental forms of astrophysical data. Light curves are modulated by, for example, stellar rotation \citep[e.g.,][]{Castro-Segura2025, Castro-Segura2025b} and/or binary orbits \citep[e.g.,][]{Alvarez-Hernandez2021}, exo-planet transits \citep[e.g.,][]{Borucki1984}, magnetic activity \citep[e.g.,][]{S-M2016}, and stellar pulsations \citep[e.g.,][]{Kilkenny1997, Green2003}. In some cases, light curves can also capture {\it transient} phenomena such as various types of novae/outbursts (e.g., dwarf nova outbursts; \citealt{Osaki1996}, micronovae; \citealt{Scaringi2022}, or supernovae; \citealt{Jha2019}).

When analysing a light curve, the inferences that can be made depend upon the type of system that was observed and the quality of the light curve. Most directly, the time-scales and amplitudes of stellar spin, binary orbits, and/or quasi-periodic oscillations \citep[QPOs; e.g.,][]{Veresvarska2024} can all be measured from the light curve. In less direct cases, parameters inferred from the light curve may be relatable to physical properties of the system. In the context of exo-planet transits, for example, the transit depth can be related to the exo-planet's radius \citep[e.g.,][]{Charbonneau2007}. As another example, Type Ia supernovae are interesting as ``standardizable candles'': the peak brightness, duration, and colour of a Type Ia supernova can be used to infer its cosmological distance \citep[e.g.,][]{Jha2019}. Moreover, high time-resolution observations of a Type Ia supernova's rise may provide insights into the nature of the progenitor \citep[e.g.,][and references therein]{Deckers2022}. It is therefore clear that high-quality light curves are vital to scientific inference.

With modern digital telescopes, constructing light curves requires software. Since telescopes generally observe and operate differently, tailored reduction software is often needed to efficiently interact with a particular instrument's data products (see the Space Telescope Science Institute's software suite,\footnote{\url{https://github.com/orgs/spacetelescope}} for example). Tailored reduction software can also account for a given instrument's ``quirks'', such as known systematic artefacts that may be confused for intrinsic variability \citep[e.g.,][]{Kinemuchi2012}. A further benefit of dedicated reduction software is that it enables reproducible reduction.

Before light curves can be constructed, however, there are a number of standard corrections that need to be applied \citep{Howell2006}. Firstly, the flux contribution due to thermal noise in the detector needs to be quantified and removed. Secondly, imperfections in the detector and optics need to be measured and accounted for. Additional systematic calibrations may be required depending on the detector. For example, the bias, a positive offset that prevents pixels from reporting negative values, artificially increases the noise floor of the detector. Beyond systematic effects, there are also observational effects that need to be corrected for. Most notably, science images will contain background contributions that will generally vary both across an image and over time; for ground-based observatories, atmospheric turbulence also introduces additional variability. For details on each of these standard corrections, see, for example, \cite{Howell2006}.

In addition to the various corrections that need to be applied, the type of photometry to perform may also warrant consideration. For both extended and point sources, it is simplest, yet still quite effective, to perform aperture photometry \citep[as in, for example, \texttt{Source-Extractor};][]{Bertin1996}. Aperture photometry involves centring an aperture of a suitable size and shape (generally circular or elliptical) on a source within an image, and then integrating the flux within the aperture to measure the source's flux. For point sources, \cite{Howell1989} showed that the highest signal-to-noise ratio (S/N) is achieved when the aperture radius is approximately equal to the full-width at half-maximum (FWHM) of the instrument's point spread function (PSF). Further improvements to the S/N may be achieved for point sources using profile-fitting photometry, provided the PSF of the instrument is well characterised. Profile-fitting photometry involves constructing either an analytical or empirical PSF model, fitting this model to a point source within an image, and then integrating the model to measure the point source's flux \citep[see, for example, Chapter 5.2 of][and references therein]{Howell2006}. Profile-fitting photometry also has the advantage of reducing contamination in crowded fields compared to aperture photometry. However, profile-fitting photometry can lead to brightness-dependent biases if the true PSF deviates considerably from the model. A more robust alternative to profile-fitting photometry, that comes with many of the same benefits, is the optimal extraction algorithm proposed by \cite{Naylor1998}; we refer to this as ``optimal photometry''. To perform optimal photometry, an assumed PSF model is centred on a point source within an image and the image pixels are weighted by this PSF model. Additionally, each pixel's weight is inversely proportional to its variance, mitigating the brightness-dependent bias incurred by profile-fitting photometry. The ``optimal flux'' is then computed by integrating the {\it weighted flux} of the entire science image. For faint sources, \cite{Naylor1998} showed that optimal photometry can improve the signal-to-noise ratio (S/N) by up to $\sim 10$ per cent compared to aperture photometry.

\subsection{OPTICAM}\label{sec: OPTICAM intro}

The OPtical TIming CAMera (OPTICAM) is a triple-camera system mounted on the 2.1~m telescope at the Observatorio Astron\'omico Nacional in San Pedro M\'artir (OAN-SPM), M\'exico. OPTICAM consists of three Andor Zyla 4.2-Plus scientific complementary metal-oxide-semiconductor (sCMOS) cameras and a set of SDSS $ugriz$ filters, enabling simultaneous three-colour observations down to sub-second time-scales \citep{Castro2019,Castro2024b,Castro2024}. OPTICAM's cameras support 2$\times$2, 3$\times$3, 4$\times$4 and 8$\times$8 on-camera pixel binning. Pixel binning can significantly reduce data transfer and storage requirements. However, on-camera binning increases the risk of saturation. Due to this saturation risk, the 8$\times$8 binning mode is not supported by OPTICAM. The pixel scales of OPTICAM's cameras are 0.1397, 0.1406, and 0.1661 arcsec/pixel for Cameras 1 ($u$-/$g$-bands), 2 ($r$-band), and 3 ($i$-/$z$-bands), respectively, while the average seeing at the OAN-SPM is $\sim 0.6$ arcsec\footnote{Within OPTICAM's telescope dome, however, the effective seeing is typically $> 1$ arcsec.} \citep{Echevarria1998, Michel2003}. For a full description of OPTICAM's hardware, see Table 5 of \cite{Castro2019}.

The sCMOS detectors used by OPTICAM differ in a number of ways from the charge-coupled device (CCD) detectors more commonly used by existing instruments \citep[e.g., ULTRACAM/HiPERCAM;][]{Dhillon2007, Dhillon2021}. Most notably, sCMOS detectors have dedicated readouts for each pixel; this has advantage of eliminating artefacts commonly seen in CCDs like hot/dark columns, saturation bleeding, etc., significantly increasing the number of usable pixels \citep[e.g.,][]{Karpov2019}. Additionally, sCMOS detectors offer faster readout speeds, lower read noise, and comparable quantum efficiency and linearity \citep[e.g.,][]{Khandelwal2024, Apergis2025}. However, due to each pixel having its own dedicated circuitry, sCMOS detectors typically have worse uniformity and stability compared to CCDs \citep[e.g.,][]{Karpov2019, Apergis2025}. In the case of OPTICAM, for example, its detectors are known to generate randomly-varying ``warm'' pixels for exposure times longer than 10~s \citep{Paez2026}. Furthermore, sCMOS detectors feature additional/increased sources of noise (e.g., amplifier glow; \citealt{Apergis2025}, random telegraph signal; \citealt{Wang2006}, row noise; \citealt{Shao2024}), and typically have higher dark currents \citep[e.g.,][]{Khandelwal2024, Apergis2025}.

When observing with OPTICAM, the observer must select each camera's filter, the pixel binning, and the exposure time. The filters used depend on the science case, the exposure time depends on both the science case and the brightness of the target source, and the pixel binning is largely determined by the brightness of the target/comparison sources. Given the advantages of pixel binning discussed above, it is generally desirable to use the largest pixel binning possible; while pixel binning reduces image resolution, this is not necessarily a problem for photometry (provided sources can still be resolved).

In this work, we present \phoptic{}, an open source photometry pipeline written in Python. \phoptic{} began as a piece of tailored software for the OPTICAM instrument. By developing a flexible instrument interface, however, it is now straightforward to use \phoptic{} to reduce standard-format FITS images produced by any imaging instrument. \phoptic{}'s source code is publicly available in an online repository\footnote{\url{https://github.com/OPTICAM-instrument/phoptic}} \textbf{(DOI: 10.5281/zenodo.21371821)} to freely download, modify, and use, and documentation is available on Read the Docs\footnote{\url{https://phoptic.readthedocs.io/en/latest}}. In Section \ref{sec: data reduction}, we describe the data reduction routines available in \phoptic{} and explain the default behaviour for each routine. In particular, in Section \ref{sec: corrections}, we describe the systematic corrections currently supported by \phoptic{}; in Section \ref{sec: background}, we describe how \phoptic{} computes image backgrounds; in Section \ref{sec: source identification}, we describe how \phoptic{} identifies sources; in Section \ref{sec: source catalogues}, we discuss how \phoptic{} constructs source catalogues; in Section \ref{sec: visualisation}, we describe some of the data quality visualisation routines currently available in \phoptic{}; finally, in Section \ref{sec: photometry}, we explain how \phoptic{} performs photometry. In Section \ref{sec: data analysis}, we detail the ``quick-look'' data analysis routines available in \phoptic{}. In Section \ref{sec: instruments}, we describe \phoptic{}'s instrument interface. In Section \ref{sec: compatibility}, we discuss how \phoptic{}'s data products can be exported to popular astronomical software. In Section \ref{sec: example}, we provide examples of \phoptic{}'s data reduction process using data from OPTICAM, ULTRACAM, HiPERCAM, and MEXMAN. Finally, we review the performance scaling of \phoptic{} in Section \ref{sec: performance}.

\section{Data reduction}\label{sec: data reduction}

The \texttt{Reducer} class in \phoptic{} provides an interface to data reduction. To instantiate \texttt{Reducer}, users must specify an output directory, for storing output files, and a data directory, containing the raw data files. When instantiating \texttt{Reducer}, \phoptic{} scans the data directory to compile an extension-agnostic list of FITS files. To interact with FITS files, \phoptic{} uses \texttt{astropy.io.fits}; as such, compressed and multi-extension FITS files are supported seamlessly.

When instantiating a \texttt{Reducer}, users can set a number of optional reduction parameters; these parameters are listed in Table \ref{tab: reducer params}, along with their corresponding default values. We explicitly note that \phoptic{}'s default parameter values were chosen with OPTICAM data in mind. For reproducibility, the parameters of the \texttt{Reducer} instance are logged to a JSON file in the output directory. During instantiation, the data are validated to ensure that all images use the same binning, and that the data are compatible with the specified \texttt{Instrument} instance (discussed in Section \ref{sec: instruments}). A design choice of \phoptic{} is that it should only read from the data directory, and \textit{never} write to it. As such, invalid data must be resolved by the user. Since the image headers are already being scanned, \phoptic{} will also parse each image's timestamp, and, by default, apply a Barycentric correction during data validation. When using the OPTICAM \texttt{Instrument}, \texttt{OPTICAM\_MX}, timestamps are shifted to mid-exposure before any Barycentric corrections are applied. In the following sections, we discuss the parameters of \texttt{Reducer} in more detail.

\begin{table*}
    \centering
    \caption{Summary of \texttt{phoptic.Reducer}'s input parameters and their default values.}
    \begin{tabularx}{\textwidth}{cX}
        \hline
        Parameter & Description \\
        \hline
        \texttt{aperture\_selector} & The function to use to compute the average PSF parameters. By default, \texttt{numpy.median} is used.\\
        \texttt{background} & The background estimator. The default background estimator uses the \texttt{photutils.background.Background2D} class with its default values.\\
        \texttt{barycenter} & Whether to apply Barycentric corrections to the image timestamps, by default \texttt{True}. Requires coordinate information in the image headers, which was not included by OPTICAM before 2023.\\
        \texttt{bias\_corrector} & The \texttt{phoptic.BiasCorrector} instance to use for applying bias corrections. If undefined, no bias corrections are applied.\\
        \texttt{dark\_corrector} & The \texttt{phoptic.DarkNoiseCorrector} instance to use for applying dark noise corrections. If undefined, no dark current corrections are applied.\\
        \texttt{finder} & The source finder routine. By default, \texttt{photutils.segmentation.SourceFinder} is used with its default values.\\
        \texttt{flat\_corrector} & The \texttt{phoptic.FlatFieldCorrector} instance to use for applying flat field corrections. If undefined, no flat-field corrections are applied.\\
        \texttt{image\_filter} & The convolution kernel/filter to apply to images as they are loaded into memory. If undefined, no filter is applied.\\
        \texttt{instrument} & The \texttt{Instrument} instance that allows \phoptic{} to interact with the data products of a particular instrument. By default, \texttt{phoptic.instruments.OPTICAM\_MX()} is used, which is the interface to OPTICAM.\\
        \texttt{rebin\_factor} & Factor by which to re-bin images, by default 1 (i.e., no rebinning). Must be an integer $\geq 1$.\\
        \texttt{remove\_cosmic\_rays} & Whether to remove cosmic rays using the L.A.Cosmic algorithm \citep{vanDokkum2001} as implemented in \texttt{astroscrappy} \citep{McCully2018}, by default \texttt{False}.\\
        \texttt{threshold} & The source detection threshold in units of background RMS, by default 5.\\
        \hline
    \end{tabularx}
    \label{tab: reducer params}
\end{table*}

\subsection{Minimal workflow}\label{sec: minimal workflow}

\begin{figure*}
    \centering
    \includegraphics[width=\textwidth]{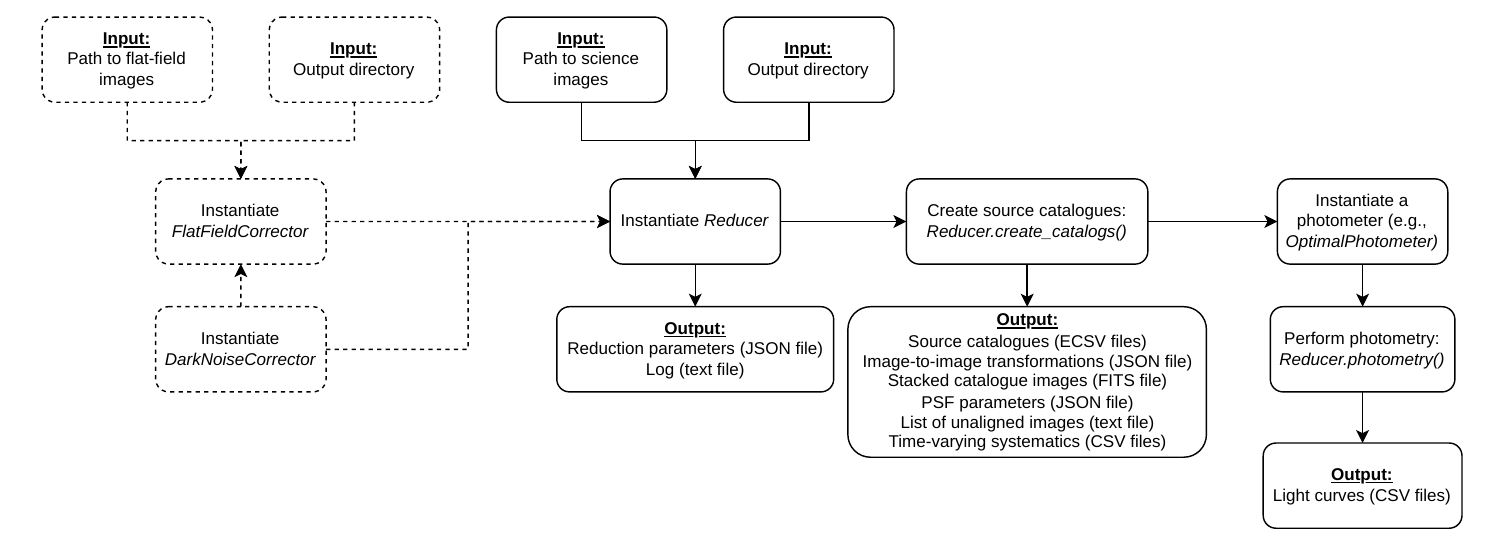}
    \caption{Minimal workflow for reducing OPTICAM data with \phoptic{}. Dashed boxes represent optional, but highly recommended, steps.}
    \label{fig: reduction flowchart}
\end{figure*}

In Figure \ref{fig: reduction flowchart}, we present a flowchart showing the minimal workflow for reducing OPTICAM data with \phoptic{}. To perform dark noise and flat-field corrections (strongly recommended), the respective classes must be instantiated (discussed in more detail in Section \ref{sec: corrections}). Bias corrections are nominally not required for OPTICAM (see Section \ref{sec: bias corrections}).

\subsection{Corrections}\label{sec: corrections}

As mentioned in Section \ref{sec: intro}, several standard corrections need to be applied during the reduction process. To perform systematic corrections, \phoptic{} includes a \texttt{correctors} module that implements dedicated routines for performing bias, dark current, and flat-field corrections. Each correction routine takes an output directory path, for storing the resulting master calibration images, and a directory path to some raw calibration images. The raw calibration images are validated to ensure they were all taken in the same binning mode, and some correction-specific validations are also performed (discussed in more detail below). Once created, master calibration images are saved to the specified output directory in FITS format. \phoptic{} also automatically propagates the errors introduced by these corrections when they are applied, including when calibration images are themselves calibrated. For example, master flat-field images should be constructed from bias- and dark noise-subtracted flat-field images, and the resulting error in the master flat-field image must account for these additional error contributions. \phoptic{}'s full photometric error propagation is discussed in Section \ref{sec: error propagation}.

\subsubsection{Bias corrections}\label{sec: bias corrections}

In the case of OPTICAM, bias corrections are performed by the cameras in real-time \citep{Castro2019}, and so manual bias corrections are nominally not required. For compatibility with other instruments, however, \phoptic{} includes a \texttt{BiasCorrector} class that can be used to perform bias corrections using the mean bias image for each camera. When a \texttt{BiasCorrector} is instantiated, the specified bias images are scanned to ensure they all have the same exposure time; since some instruments may not support 0~s exposures, \phoptic{} does not enforce a 0~s exposure time for bias images. If a \texttt{BiasCorrector} instance is passed to \texttt{Reducer}, then this is the first correction applied to all images.

\subsubsection{Dark noise corrections}\label{sec: dark noise corrections}

After subtracting the bias, the next calibration that needs to be applied is the dark current or dark noise correction. In \phoptic{}, dark noise is corrected by the \texttt{DarkNoiseCorrector} class. Unlike \texttt{BiasCorrector}, \texttt{DarkNoiseCorrector} can be instantiated without any input images, in which case \phoptic{} will attempt to infer an image's dark noise using the exposure-integrated dark current. However, this requires that science images report the detector's dark current in their image headers. If no dark current is reported, dark images may be used instead.

Since dark images require a non-zero exposure time, and may therefore include cosmic rays, \phoptic{} applies dark noise corrections using the median dark image for each camera. Additionally, dark images will include a bias contribution, and so a \texttt{BiasCorrector} instance may be passed to \texttt{DarkNoiseCorrector} to automatically bias-correct the dark images. Finally, before dark noise corrections are applied, \phoptic{} will check that the exposure times of the dark images match those of the science images.

\subsubsection{Flat-field corrections}\label{sec: flat corrections}

The final calibration performed by \phoptic{} is flat-fielding, which is handled by the \texttt{FlatFieldCorrector} class. To bias- and dark-correct flat-field images, \texttt{BiasCorrector} and \texttt{DarkNoiseCorrector} instances can be passed to \texttt{FlatFieldCorrector}. As mentioned in Section \ref{sec: dark noise corrections}, the exposure times of the dark images must match the exposure times of the ``science images''; when passing a \texttt{DarkNoiseCorrector} instance to \texttt{FlatFieldCorrector}, the flat-field images are considered science images. Similarly to \texttt{DarkNoiseCorrector}, \texttt{FlatFieldCorrector} computes the master flat-field image using the median flat-field image for each camera--filter combination.

\subsection{Convolution kernels/filters}\label{sec: image filters}

As mentioned in Section \ref{sec: OPTICAM intro}, \cite{Paez2026} recently found that OPTICAM's cameras generate randomly-varying warm pixels during long exposures ($\geq 10$~s). To mitigate this, \cite{Paez2026} suggest applying a 3$\times$3 median filter to calibration and science frames before reduction. For generality, \phoptic{} extends this to allow for the specification of arbitrary convolution kernels/filters via the \texttt{image\_filter} parameter of the \texttt{BiasCorrector}, \texttt{DarkNoiseCorrector}, \texttt{FlatFieldCorrector}, and \texttt{Reducer} classes. We note, however, that applying a kernel/filter likely introduces correlated systematics between pixels that are not accounted for in \phoptic{}'s error propagation (Section \ref{sec: error propagation}).

\subsection{Software pixel binning}\label{sec: software rebinning}

On-the-fly software pixel binning can be performed by passing an integer greater than 1 to the \texttt{rebin\_factor} parameter of the \texttt{BiasCorrector}, \texttt{DarkNoiseCorrector}, \texttt{FlatFieldCorrector}, and/or \texttt{Reducer} classes. Unlike on-camera pixel binning, software pixel binning does not incur an increased risk of pixel saturation. However, since this re-binning is done on-the-fly, the performance benefits are much less pronounced than those of on-camera pixel binning. On-the-fly software pixel binning is primarily included for cases in which there is a mismatch in resolution between calibration and science images, and/or a source was observed using multiple binning modes - in which case it may be desirable to re-bin all images to a common resolution.

\subsection{Background estimation}\label{sec: background}

\begin{figure}
    \centering
    \includegraphics[width=\columnwidth]{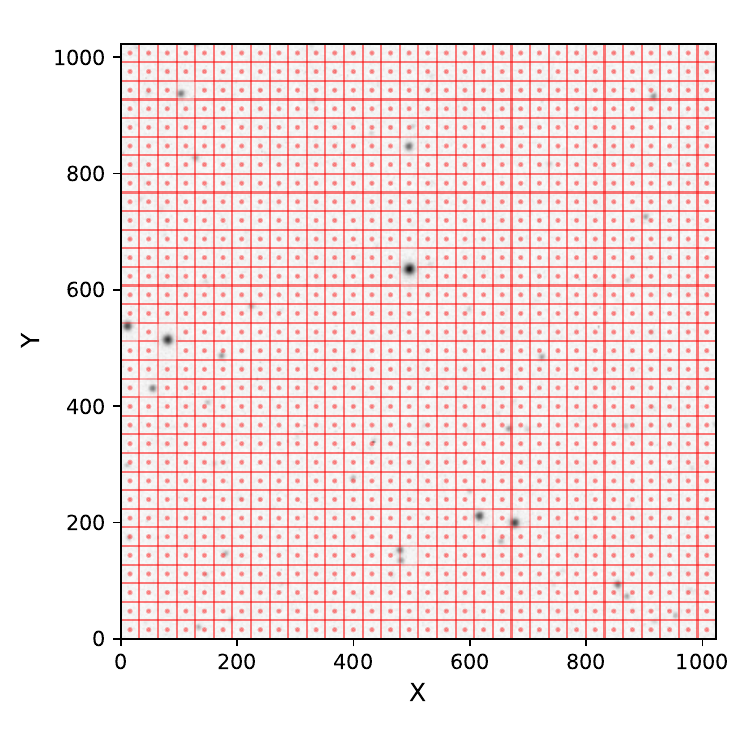}
    \caption{\phoptic's default background mesh (red) for an \textbf{OPTICAM} image taken in 2$\times$2 binning mode. Mesh boxes used to estimate the background are marked with red dots.}
    \label{fig: background mesh}
\end{figure}

\phoptic{} uses the \texttt{photutils.background} package to compute two-dimensional backgrounds for each image \citep{photutils}. The default behaviour is to compute an image's background using \texttt{photutils.background.Background2D} with the \texttt{box\_size} parameter set to the image width divided by 32, and all other parameters left at their default values. This divides the image into a 32$\times$32 mesh, and estimates the average background and its root-mean-square (RMS) within each box. By default, \texttt{photutils.background.Background2D} uses iterative sigma clipping, with a threshold of $3 \sigma$ from the median and a maximum of 10 iterations, to clip anomalous pixels and/or entire mesh boxes from the background estimation. The background and its corresponding RMS in excluded mesh boxes is then estimated by interpolating the estimates of neighbouring boxes. An example of an image taken using the 2$\times$2 binning mode and its corresponding background mesh is shown in Figure \ref{fig: background mesh}. Mesh boxes used to compute the background image are identified with red dots.

To estimate the average background within a box, \texttt{photutils.background.Background2D} defaults to the \texttt{Source-Extractor} mode estimator:
\begin{equation}
    \text{background} = 
    \begin{cases}
        2.5 \tilde{f} - 1.5 \bar{f} & \text{if ($\bar{f}$ - $\tilde{f}$) / $\sigma_f$ > 0.3} \\
        \tilde{f}                   & \text{otherwise},
    \end{cases}
    \label{eq: SExtractor background}
\end{equation}
where $\tilde{f}$ and $\bar{f}$ represent the median and mean flux within the box, respectively, and $\sigma_f$ represents the standard deviation of the flux within the box \citep{Bertin1996}. To estimate the background RMS, \texttt{photutils.background.Background2D} defaults to the standard deviation \citep{photutils}. By default, \phoptic{} should therefore produce similar background estimates to \texttt{Source-Extractor}. However, \texttt{photutils.background} provides several background and background RMS estimators, and allows users to specify custom routines; as such, background estimation with \phoptic{} can be tailored if this default is inadequate.

\subsection{Source identification}\label{sec: source identification}

To identify astronomical sources, \phoptic{} uses the \texttt{photutils.segmentation} module. By default, \phoptic{} identifies sources using the \texttt{photutils.segmentation.SourceFinder} routine, which combines source detection and deblending. \texttt{SourceFinder} requires manually specifying an \texttt{n\_pixels} parameter, which quantifies how many connected pixels need to be above a specified threshold to constitute a source. In \phoptic{}, \texttt{n\_pixels} defaults to:
\begin{equation}
    {\tt n\_pixels} = \frac{128}{b_f^2},
\end{equation}
where $b_f$ is the binning factor. The binning factor accounts for the hardware pixel binning used during observation and the {\tt rebin\_factor} parameter of the \texttt{Reducer} instance. For example, if the observation used the ``2$\times$2'' hardware pixel binning mode and the \texttt{Reducer} instance was defined with \texttt{rebin\_factor=2}, then the binning factor would be $b_f = 2 \times 2 = 4$. The source detection threshold is determined by the \texttt{Reducer} instance's \texttt{threshold} parameter; this is the factor above the background RMS that \texttt{n\_pixels} connected pixels must be to be considered a source. By default, \texttt{threshold} is set to 5. All other \texttt{photutils.segmentation.SourceFinder} parameters are left to their default values.

\begin{figure}
    \centering
    \includegraphics[width=\columnwidth]{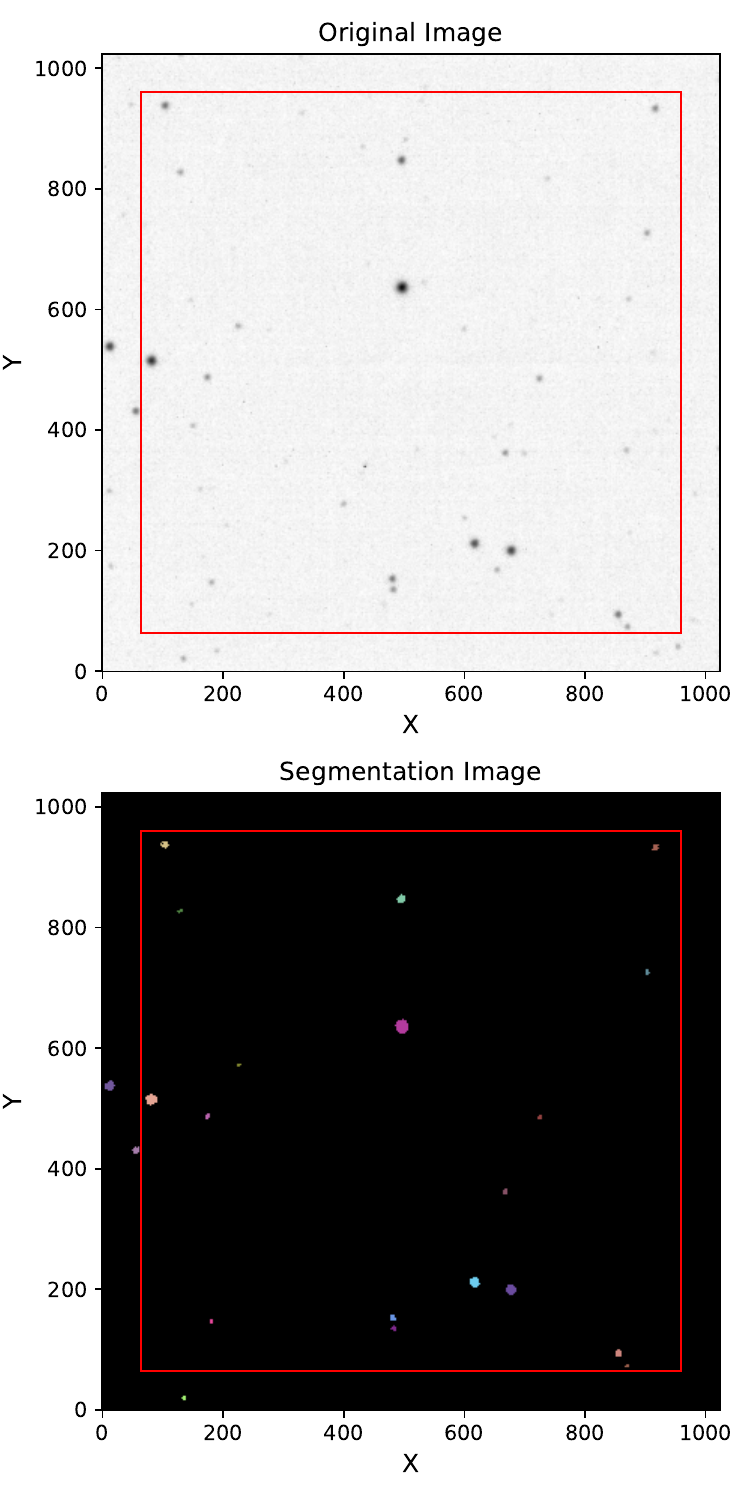}
    \caption{Top: OPTICAM $g$-band image taken in 2$\times$2 binning mode; bottom: corresponding segmentation image computed using \phoptic's default source finder. The coloured segments in the bottom image represent the footprints of detected sources. The red square in both images represents the exclusion border for catalogue sources (see Section \ref{sec: source identification} for more details).}
    \label{fig: segmentation image}
\end{figure}

When called, \texttt{photutils.segmentation.SourceFinder} returns a \texttt{photutils.segmentation.SegmentationImage} instance. Figure \ref{fig: segmentation image} shows an example of a segmentation image, as well as the image from which it was computed. An additional feature of \phoptic{}'s default source identification routine is that sources close to the edge of an image are excluded. In Figure \ref{fig: segmentation image}, the red boxes show this source exclusion border, which corresponds to 1/16th of the image's width. This exclusion border is used to prevent partial sources from being detected.

In \phoptic{}'s default source identification routine, the \texttt{SegmentationImage} instance returned by \texttt{SourceFinder} is converted to an \texttt{astropy.table.QTable} instance. This provides a human-readable representation of the source catalogue, and enables easy reading/writing of source catalogues from/to disk. If the default behaviour of either \phoptic{} or \texttt{SourceFinder} is poorly-suited to a particular observation, users may pass a custom routine to the \texttt{finder} parameter of \phoptic's \texttt{Reducer} class.

\subsection{Creating source catalogues}\label{sec: source catalogues}

\begin{figure*}
    \centering
    \includegraphics[width=\textwidth]{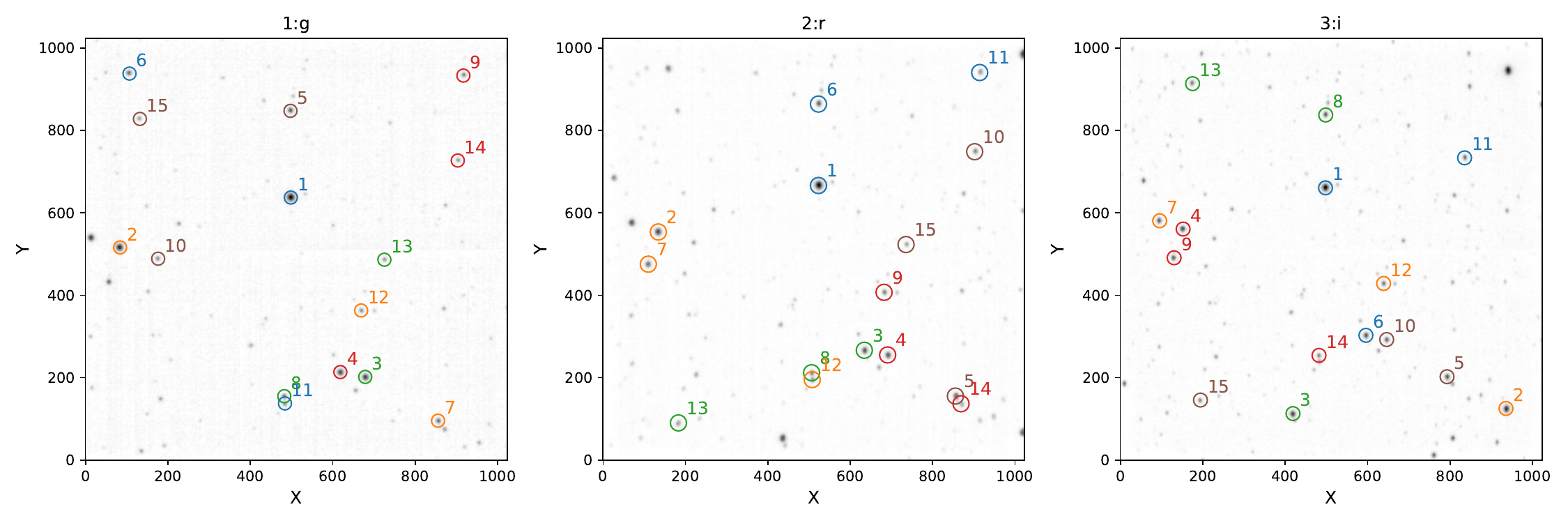}
    \caption{An example of a \phoptic{} source catalogue \textbf{constructed from an OPTICAM observation}. For this observation, the 2$\times$2 binning mode was used, Camera 1 used the $g$-band filter, Camera 2 used the $r$-band filter, and Camera 3 used the $i$-band filter.}
    \label{fig: source catalogue}
\end{figure*}

The \texttt{create\_catalogs()} method of \phoptic's \texttt{Reducer} class aligns and stacks each camera's images (improving source S/N), identifies sources in the stacked images, and then creates a source catalogue for each camera. Source IDs within each catalogue are determined by source brightness, with the brightest source corresponding to Source 1, Source 2 being the second brightest, and so on. Due to the different FoVs and pixel-scales of OPTICAM's cameras, in addition to each camera observing in a different wavelength range, a given source will generally be assigned a different ID in each camera's catalogue. An example of a source catalogue is shown in Figure \ref{fig: source catalogue}.

When calling \texttt{create\_catalogs()}, the following parameters can be configured by the user: the maximum number of sources per catalogue (\texttt{max\_catalog\_sources}), the (maximum) number of sources to use for image alignment (\texttt{n\_alignment\_sources}), the type of alignment to perform (\texttt{transform\_type}), and the alignment transformation limits (\texttt{rotation\_limit}, \texttt{translation\_limit}, and \texttt{scale\_limit}). By default, the number of sources per catalogue and the maximum number of sources to use for image alignment are both set to 15, and no alignment limits are imposed. 

The 2.1~m telescope on which OPTICAM is mounted is equipped with a guiding system that keeps the telescope on target. As such, image-to-image alignments are nominally not required for OPTICAM. For generality, however, \phoptic{} defaults to \texttt{transform\_type=``affine"}, which uses the \texttt{astroalign} Python package to compute the image-to-image alignments. \texttt{astroalign} can align images directly; however this uses the \texttt{sep} \citep[][]{sep} Python package to compute backgrounds and identify sources, which may produce different results to the background estimator and source identification routines of the \texttt{Reducer} instance. To avoid this, \phoptic{} first performs source detection (and background subtraction) using the \texttt{Reducer} instance's routines, and then passes the source coordinates to \texttt{astroalign.find\_transform()} to compute the affine transformations between images. Alternatively, images can be aligned using simple $(x, y)$ translations instead by passing \texttt{transform\_type=``translation"}. An advantage of the translation transformation is that it can be computed from a single source, while affine transformations require at least three sources. Translation is therefore better suited to sparse fields. Additionally, the affine and translation alignment methods differ in how they treat the \texttt{n\_alignment\_sources} parameter: when computing affine transformations, the number of alignment sources is an {\it upper limit}, while translation transformations are computed using all available alignment sources.

Once source catalogues have been created, they are saved to disk using \texttt{astropy}'s Enhanced Character Separated Values (ECSV) format \citep{ECSV}. Additionally, the stacked images are saved to a compressed FITS file, the image-to-image transformations are saved to a JSON file, and the names of any files that could not be aligned are saved to a text file. When creating source catalogues, the PSF of each camera is also modelled as a 2D Gaussian. The average semi-major and semi-minor standard deviations of each camera's PSF are logged to a JSON file in both pixels and arcseconds, and the orientation of the PSF is logged in degrees (though this may vary with position on the detector; \citealt{Howell1996}). Source catalogues therefore only need to be computed once, and can be subsequently read from file.

\subsubsection{Manually identifying sources}\label{sec: source picking}

In some cases, a source of interest may be several magnitudes fainter than other sources in the field. Such sources may therefore require low \texttt{threshold} and/or large \texttt{max\_catalog\_sources} parameter values in order to be catalogued, resulting in crowded catalogues and increasing the risk of spurious source detections. To prevent this, \phoptic{} provides a routine that allows users to manually catalogue sources interactively.

To manually catalogue sources, users must first call the \texttt{create\_catalogs()} method to compute the required image-to-image transformations. Users may then call the \texttt{pick\_sources()} method of the \texttt{Reducer} class to open an interactive plot of the source catalogues (e.g., Figure \ref{fig: source catalogue}). To make faint sources more easily visible, \texttt{pick\_sources()} takes a \texttt{percentile} normalisation parameter, similar to the percentage scalings in SAOImageDS9 \citep[][]{DS9}. By default, \texttt{percentile} is set to 99. In the resulting plot, users can left-click on sources of interest to manually catalogue them. When a user clicks on a source of interest, \phoptic{} fits the PSF model of the corresponding catalogue to a small region, determined by the \texttt{region\_size} parameter, around the clicked coordinates to refine the centroid coordinates of the source. By default, the \texttt{region\_size} parameter is set to the image width divided by 64. When fitting the PSF model, the amplitude and $(x, y)$ coordinates are free parameters, while the semi-major and semi-minor standard deviations are fixed to the values determined from the corresponding catalogue (Section \ref{sec: source catalogues}). A benefit of of this approach is that it does not require a detection threshold parameter, thereby enabling detection of faint point sources.

\subsection{Visualisation routines}\label{sec: visualisation}

After creating source catalogues, but before performing photometry, users may want to, for example, check the image-to-image alignments, visualise the PSF, and inspect growth curves (measures of source flux as a function of aperture radius). \phoptic{} provides a number of convenience routines designed to aid in performing these standard diagnoses.

\subsubsection{Time-varying systematics}\label{sec: time-varying systematics}

\begin{figure*}
    \centering
    \includegraphics[width=\textwidth]{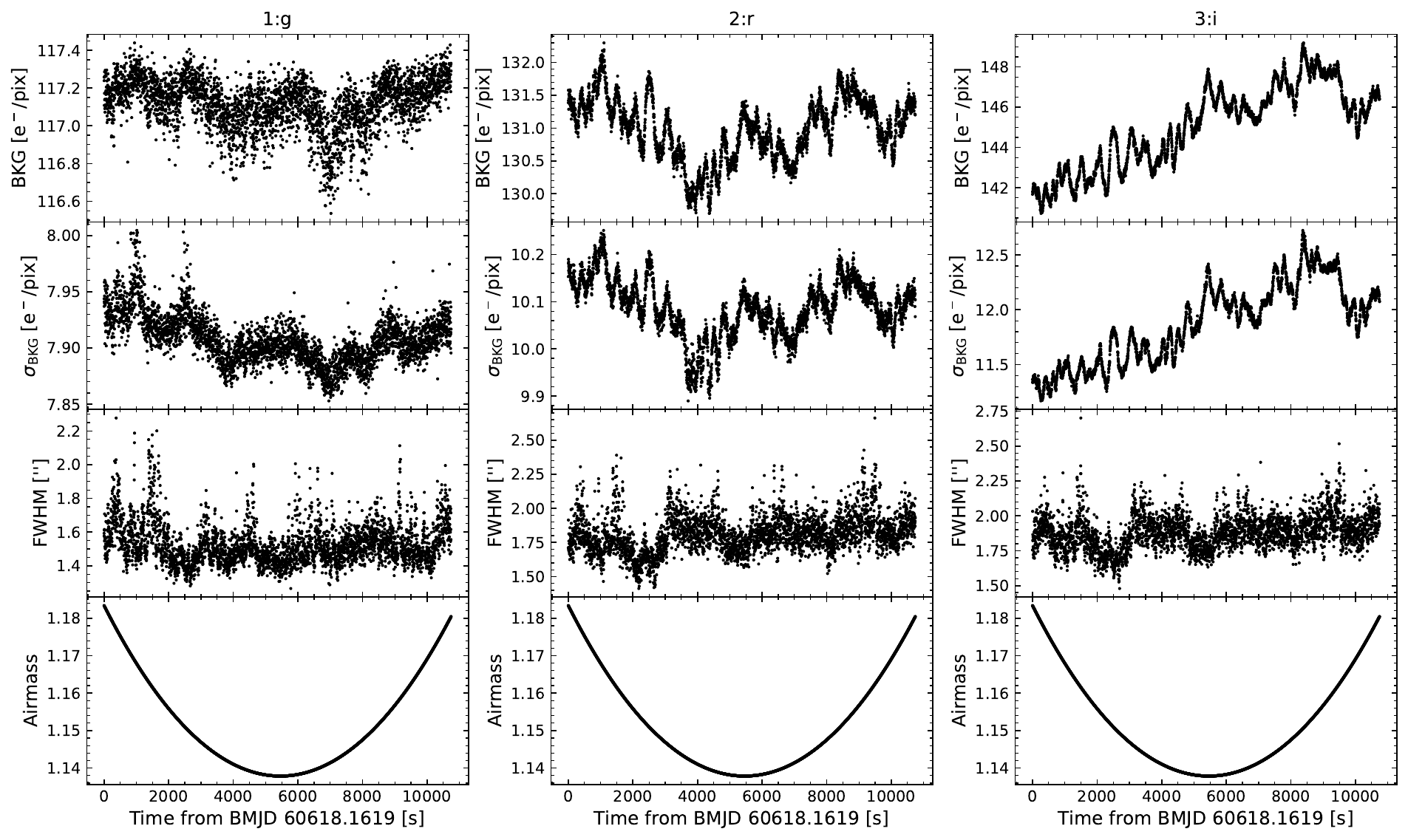}
    \caption{Time-varying systematics for the observation corresponding to Figure \ref{fig: source catalogue}.}
    \label{fig: systematics}
\end{figure*}

While computing the image-to-image alignments required to construct the source catalogues (Section \ref{sec: source catalogues}), \phoptic{} also records the average background and its RMS, the average PSF FWHM, and the airmass. These time-varying systematics are then saved to CSV files for each camera, and a plot showing how these systematics vary over time is produced; in Figure \ref{fig: systematics}, we present the systematics for the observation corresponding to Figure \ref{fig: source catalogue}.

\subsubsection{Alignment GIFs}\label{sec: alignment gifs}

To check the image-to-image alignments, users can generate alignment GIF files using the \texttt{create\_gifs()} method of the \texttt{Reducer} class. The resulting files show the catalogue source positions transformed onto each science image. Each frame of the GIF files also shows the science image's file name and extension number (for multi-extension FITS files), allowing poorly-aligned images to be identified. It may also be instructive to inspect these GIF files to verify that the alignments are suitable for performing forced photometry. If any misaligned images are identified, users can add them to the ``unaligned files'' list by passing the file names to the \texttt{update\_unaligned\_files()} method of \texttt{Reducer}.

\subsubsection{PSF visualisation}\label{sec: psf visualisation}

\begin{figure}
    \centering
    \includegraphics[width=\columnwidth]{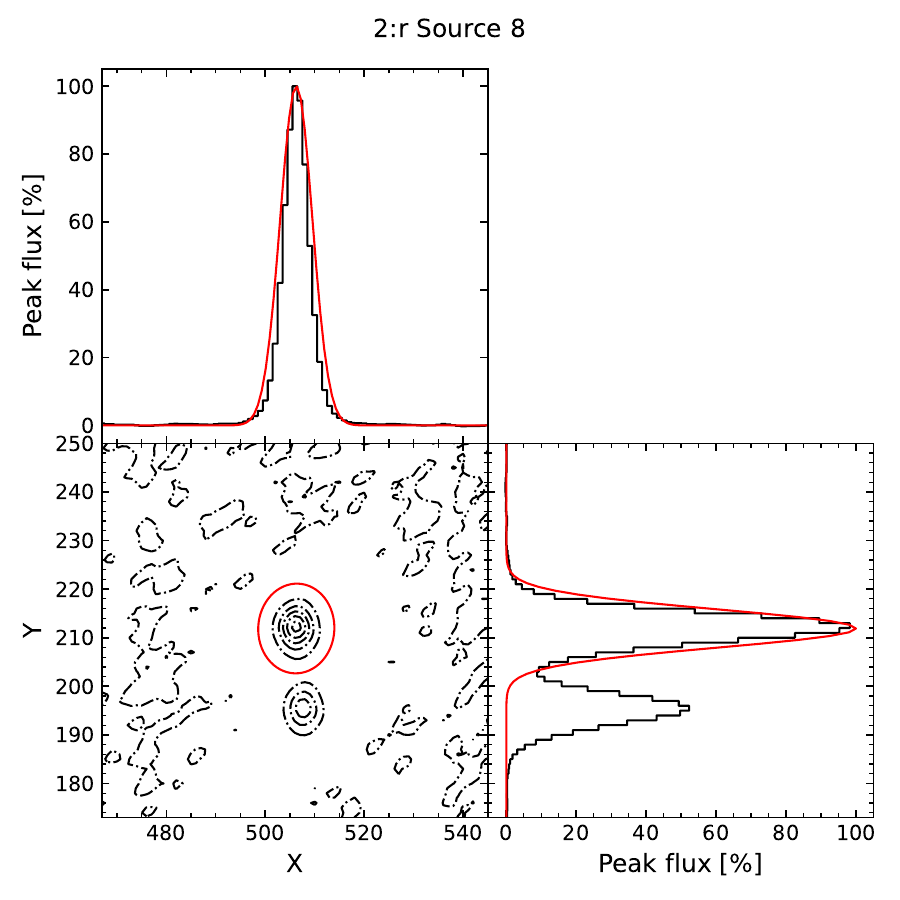}
    \caption{The PSF of Camera 2 ($r$-band) Source 8 from Figure \ref{fig: source catalogue}. Bottom left: the (stacked) catalogue image's brightness contours (black dash-dotted lines) and \phoptic's default aperture shape and size (Section \ref{sec: photometry}; solid red line). Top left: brightness profile along the y-centroid of the target source (black step function) and the average PSF projected onto the x-axis (red line). Bottom right: brightness profile along the x-centroid of the target source (black step function) and the average PSF projected onto the y-axis (red line).}
    \label{fig: PSF}
\end{figure}

An important quality check for any observation is to inspect the PSF. In \phoptic, this can be done using the \texttt{plot\_psfs()} method of the \texttt{Reducer} class. PSF visualisations generated by the \texttt{plot\_psfs()} method also compare each source's PSF to the average PSF of the corresponding camera. As an example, the PSF plot for $r$-band Source 8 from Figure \ref{fig: source catalogue}, which is in close proximity to Source 12, is shown in Figure \ref{fig: PSF}.

As can be seen from Figure \ref{fig: PSF}, OPTICAM's PSF is well-approximated by the 2D Gaussian model shown in red. Note that the PSF models shown in the top left and bottom right panels have not been fit to the brightness profiles; instead, they are the projections of the average PSF onto the x- and y-axes of the detector (accounting for the source's specific PSF orientation).

\subsubsection{Growth curves}\label{sec: growth curves}

Growth curves can be useful for defining suitable aperture sizes and quantifying the extent of extended sources. To generate growth curves in \phoptic, users can call the \texttt{plot\_growth\_curves()} method of the \texttt{Reducer} class. Growth curves are generated using \textit{circular} apertures of increasing radii centred on each source; by default, the maximum aperture radius is set to 10 times the average PSF semi-major standard deviation. For extended sources, the maximum aperture radius may therefore need to be increased.

\subsubsection{Signal-to-noise ratios}\label{sec: SNRs}

\begin{figure*}
    \centering
    \includegraphics[width=\textwidth]{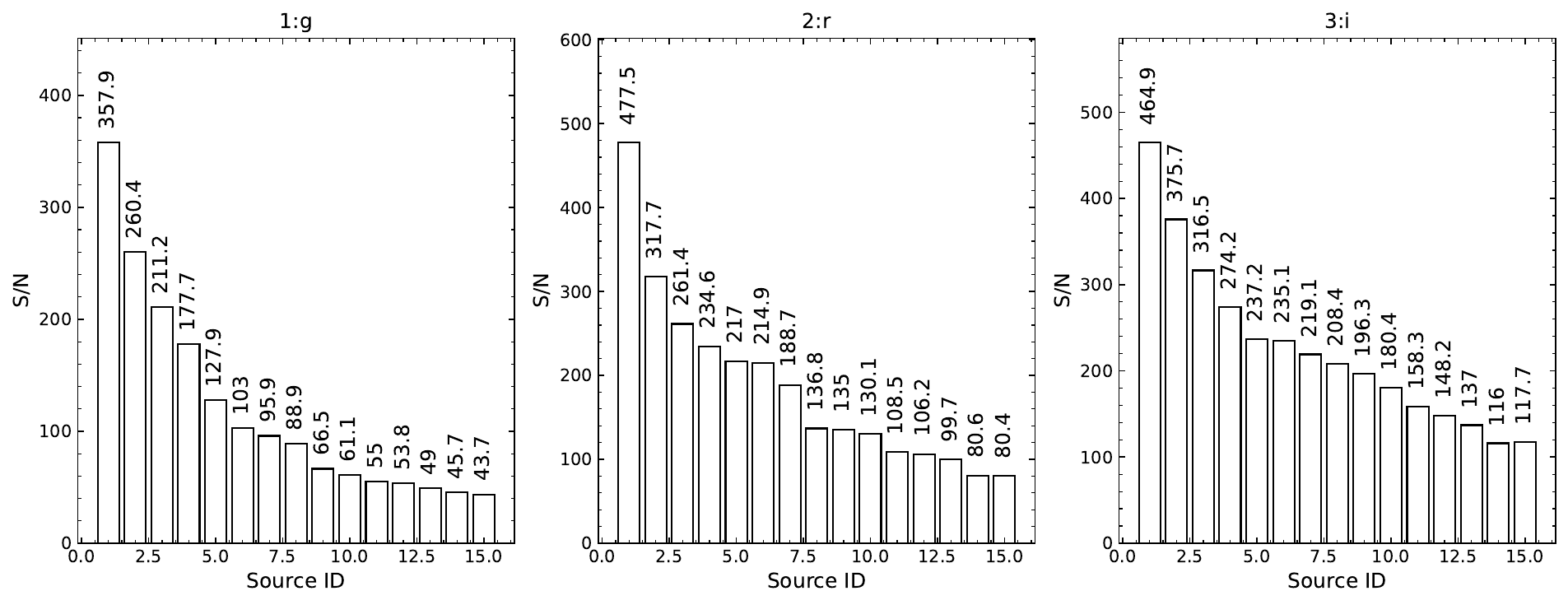}
    \caption{Corresponding S/N for the sources identified in Figure \ref{fig: source catalogue} as computed from the alignment reference images.}
    \label{fig: SNR}
\end{figure*}

To quantify the signal-to-noise ratio (S/N) of catalogued sources in \phoptic, users can call the \texttt{plot\_snrs()} method of the \texttt{Reducer} class. This computes the S/N for all catalogued sources \citep[as described in Chapter 4.4 of][]{Howell2006} using the alignment reference images; an example of which is shown in Figure \ref{fig: SNR} corresponding to the catalogues shown in Figure \ref{fig: source catalogue}. The propagated noise contributions are detailed in Section \ref{sec: error propagation}. Note, however, that the S/N for each source is computed from a single exposure, and so this may give misleading results for variable sources.

\subsection{Performing photometry}\label{sec: photometry}

After source catalogues have been constructed, users can perform photometry by defining a photometer and passing it to their \texttt{Reducer} instance's \texttt{photometry()} method. Currently, \phoptic{} provides two types of photometry: aperture photometry via the \texttt{AperturePhotometer} class, and optimal photometry \citep{Naylor1998} via the \texttt{OptimalPhotometer} class. Photometry results are saved as individual CSV files for each source in each catalogue.

The \texttt{AperturePhotometer} class performs photometry using an elliptical aperture of a given size and orientation; under-the-hood, \texttt{AperturePhotometer} uses the \texttt{photutils.aperture} module. For point sources, \cite{Howell1989} showed aperture photometry yields the highest S/N when the aperture radius \textit{approximately} corresponds to the PSF's FWHM. By default, \phoptic{} scales the aperture radius to the average PSF FWHM as computed from the corresponding catalogue image. Following \cite{Howell1989}, some tuning of \phoptic{}'s default aperture radius will generally be required to optimise the S/N for point sources. Alternatively, users may specify a custom aperture (i.e., for extended sources). We note that, once specified, the aperture size is fixed for all sources and exposures. However, since the orientation of the PSF can vary with the source's position on the detector \citep[][]{Howell1996}, the orientation of the ellipse is set per source. A source's flux is then measured by simply summing the pixels in the aperture; for partially overlapping pixels, their fractional overlap is used to scale their contribution to the measured flux.

The \texttt{OptimalPhotometer} class performs the optimal extraction algorithm described in \cite{Naylor1998}. As mentioned in Section \ref{sec: intro}, optimal photometry involves weighting each pixel by a PSF model, $P$, and the pixel's variance, $\sigma^2$:
\begin{equation}
    w_{i, j} = \frac{P_{i, j} / \sigma_{i, j}^2}{\sum_{k, l \in {\rm image}} P_{k, l}^2 / \sigma^2_{k, l}}.
    \label{eq: optimal weights}
\end{equation}
By accounting for the pixel-wise variance, optimal photometry avoids the brightness-dependent biases incurred by profile-fitting photometry. Similarly to the \texttt{AperturePhotometer} class, the orientation of the PSF is allowed to vary from source to source. A source's flux is then given by the weighted sum over the entire science image:
\begin{equation}
    F_{\rm opt} = \sum_{i, j \in {\rm image}} w_{i, j} f_{*, i, j},
    \label{eq: optimal flux}
\end{equation}
where $f_*$ represents the background-subtracted flux. From equations \ref{eq: optimal weights} and \ref{eq: optimal flux}, two things are clear in general: i) the optimal flux will be much lower than the aperture flux in terms of absolute counts; ii) optimal photometry is more computationally expensive than aperture photometry. Fortunately, differential photometry mitigates the difference in absolute counts between aperture and optimal photometry (see Section \ref{sec: differential photometry}); to improve the performance of optimal photometry, \phoptic{}'s implementation is optimised using \texttt{numba}'s just-in-time compiler \citep[][]{numba}.

\subsubsection{Local background estimation}\label{sec: local background}

Both photometry classes implemented in \phoptic{} will default to the \texttt{Reducer} instance's global background estimator. Alternatively, both photometers can be configured to compute the background around sources locally using an annulus instead. \phoptic{} includes a default local background estimator via the \texttt{DefaultLocalBackground} class, which uses an elliptical annulus around a source to compute the sigma-clipped mean and standard deviation of the local background. The sigma clipping routine is identical to the global background's routine discussed in Section \ref{sec: background}. The inner and outer axes of the annulus are specified in units of the semi-major and semi-minor standard deviations of the average PSF. By default, the inner axes of the annulus are set to 5 times the average PSF standard deviations, while the outer axes of the annulus scale the standard deviations of the average PSF by a factor 7.5. Alternatively, users may pass custom routines, should they want to compute the local background differently.

\subsubsection{Aperture visualisation}\label{sec: aperture visualisation}

\begin{figure}
    \centering
    \includegraphics[width=\columnwidth]{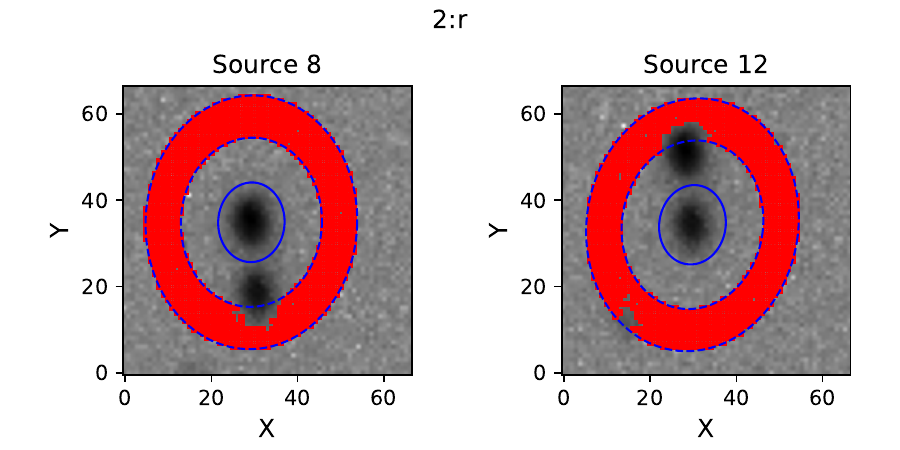}
    \caption{Visualisations of the default size and shape of the \texttt{AperturePhotometer} aperture (solid blue line) and \texttt{DefaultLocalBackground} annulus (dashed blue line) for Camera 2 ($r$-band) Sources 8 and 12 from the catalogue presented in Figure \ref{fig: source catalogue}. The red boxes represent pixels included in the local background estimation.}
    \label{fig: aperture example}
\end{figure}

When defining an aperture and/or annulus, it can be instructive to verify the shape and size of the aperture and/or annulus on the science images themselves. In \phoptic{}, this can be done by passing a photometer instance to the \texttt{plot\_apertures()} method of the \texttt{Reducer} class. If the photometer instance uses an annulus, the (sigma-clipped) pixels used to estimate the local background are also visualised. For visualisation, the alignment reference images are used. As an example, Figure \ref{fig: aperture example} shows the default aperture and annulus size for \texttt{AperturePhotometer} and \texttt{DefaultLocalBackground}, respectively, for $r$-band Sources 8 and 12 from the source catalogue shown in Figure \ref{fig: source catalogue}.

As can be seen from Figure \ref{fig: aperture example}, the default aperture appears to be of a reasonable size and shape for these two point sources. Moreover, it can be seen that the default annulus effectively clips outlier pixels from the local background estimation.

\subsubsection{Forced photometry}\label{sec: forced photometry}

By default, both photometry classes implemented in \phoptic{} will only compute the flux of a source from a given image if it is detected by the specified source finder in said image. To disable this, users can pass \texttt{forced=True} when instancing the desired photometry class. If \texttt{forced=True}, the image-to-image transformations will be used to perform photometry at the source's expected position, regardless of whether or not it is identified by the source finder (i.e., ``forced photometry''). This is particularly useful for faint sources, or sources that may exhibit drastic decreases in flux (e.g., eclipsing sources).

\subsubsection{Error propagation}\label{sec: error propagation}

When performing photometry, \phoptic{} accounts for up to seven error/noise sources: bias variance, $\sigma_{\rm bias}^2$, dark noise variance, $\sigma_{\rm dark}^2$, flat-field variance, $\sigma_{\rm flat}^2$, background variance (often referred to as sky noise), $\sigma_{\rm bkg}^2$, shot noise variance, $f_*$, read noise variance, $\sigma_{\rm rn}^2$, and scintillation noise variance, $\sigma_{\rm scint}^2$. The first three error sources are a result of variance in the calibration images. Similarly, background estimates are made over a number of pixels, introducing an additional variance term. Shot noise arises from counting statistics, and has the convenient property that its variance is equal to the measured flux, $f_*$. Read noise is introduced by reading from the detector, and its contribution is independent of both the measured flux and exposure time; in the case of OPTICAM's cameras, the nominal read noise RMS, $\sigma_{\rm rn}$, is 1.3 electrons/pixel \citep[][]{Castro2019}. The pixel-wise variance from each of these sources is therefore:
\begin{equation}
    \sigma_{i, j}^2 = [\sigma_{\rm bias, i, j}^2] + [\sigma_{\rm dark, i, j}^2] + [\sigma_{\rm flat, i, j}^2] + \sigma_{\rm bkg, i, j}^2 + \sigma_{\rm rn, i, j}^2 + f_{*, i, j},
    \label{eq: pixel-wise error}
\end{equation}
where the square brackets denote optional terms that will only be included if the respective corrector has been passed to the \texttt{Reducer} instance (Section \ref{sec: corrections}). Finally, scintillation noise results from turbulence in the atmosphere. Unlike the above noise sources, however, scintillation noise is a \textit{relative} contribution.

To quantify scintillation noise, atmospheric turbulence profiling is required \citep[][]{Osborn2015}. Alternatively, the scintillation noise can be approximated using Young's approximation \citep{Young1967}; by default, \phoptic{} implements the improved version of Young's approximation presented in \cite{Osborn2015}:
\begin{equation}
    \sigma_{\rm scint}^2 = 10^{-5} C^2 D^{-4/3} t^{-1} a^{-3} \exp(-2 h / H) F^2,
    \label{eq: scintillation noise}
\end{equation}
where $C$ is the \textit{site-dependent} empirical correction coefficient, $D$ is the telescope diameter in metres, $t$ is the exposure time in seconds, $a$ is the airmass, $h$ is the altitude of the observatory in metres, $H$ is the scale-height of atmospheric turbulence in metres (generally accepted to be $\sim 8000$~m), and $F$ is the measured source flux. At OPTICAM's observing site in San Pedro M\'artir, \cite{Osborn2015} found that $C = 1.67$. The effective noise, $\sigma_{\rm eff}$, when performing aperture photometry is then given by:
\begin{equation}
    \sigma_{\rm eff} = \sqrt{\sigma_{\rm scint}^2 + \sum_{i, j \in {\rm aperture}} \sigma_{i, j}^2}.
    \label{eq: total error}
\end{equation}
In the case of optimal photometry, the aperture sum in equation \ref{eq: total error} is replaced with a \textit{weighted} sum over the entire science image.

\begin{figure*}
    \centering
    \includegraphics[width=\textwidth]{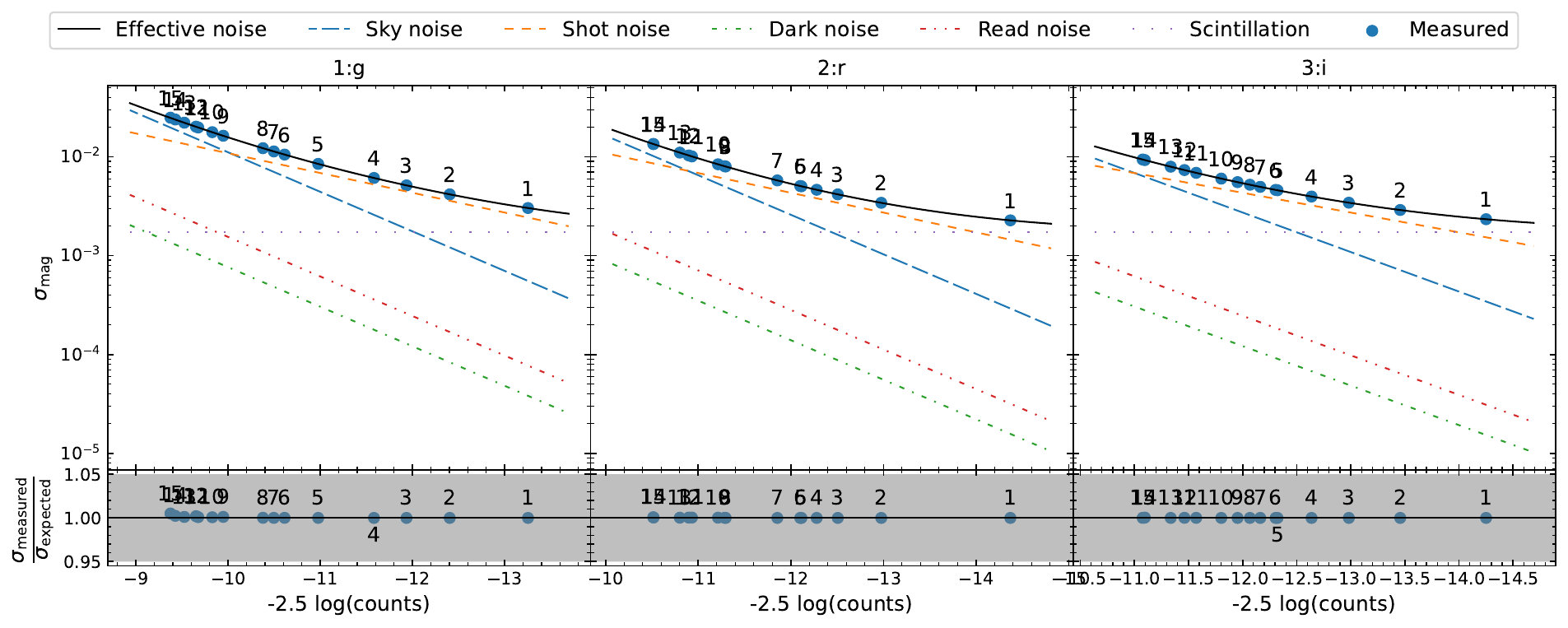}
    \caption{Measured flux error and noise contributions for the catalogued sources shown in Figure \ref{fig: source catalogue}. The solid line shows the total (i.e., effective) noise, the long dashed line represents the sky/background noise, the short dashed line represents the shot noise, the dashdotted line represents the nominal dark noise, the dashdotdotted line represents the nominal read noise, and the loosely dotted line represents the scintillation noise.}
    \label{fig: noise characterisation}
\end{figure*}

To visualise the contributions from the above noise sources, noise characterisation plots can be generated. To generate these plots, \phoptic{} performs aperture photometry using the default aperture size (Section \ref{sec: photometry}). As an example, Figure \ref{fig: noise characterisation} presents the noise characterisation plot corresponding to the catalogue shown in Figure \ref{fig: source catalogue}. We recall that OPTICAM's cameras perform automatic bias corrections in real-time, and so the error introduced by this correction cannot be quantified. Additionally, we have neglected flat-field corrections in this case.

As can be seen from Figure \ref{fig: noise characterisation}, many of the sources identified in Figure \ref{fig: source catalogue} are shot noise dominated, while the brightest sources are scintillation noise dominated. For this particular observation, the exposure time was 3~s, and it can be seen that the dark noise contribution is approximately half that of the nominal read noise contribution. Deviations of the measured noise from the effective noise indicate spatially-varying noise contributions. In the case of Figure \ref{fig: noise characterisation}, it can be seen that spatial variations in the noise contributions are less than one per cent.

\subsubsection{Comparison with \texttt{Source-Extractor}}\label{sec: SExtractor comparison}

\texttt{Source-Extractor} \citep{Bertin1996} is a popular astronomical image reduction software that has been used in the reduction of data from a number of instruments, including the DESI Legacy Imaging Surveys \citep[e.g.,][]{Dey2019}, the Galaxy Evolution Explorer (GALEX; \citealt{Morrissey2007}), Pan-STARRS \citep[e.g.,][]{Scolnic2018}, and the Zwicky Transient Facility (ZTF; \citealt{Bellm2019, Masci2019}). It is therefore instructive to compare \phoptic{}'s results to those obtained by \texttt{Source-Extractor} when both are configured similarly. To do this, we used both programs to perform aperture photometry on the observation represented in Figure \ref{fig: source catalogue}. We used \phoptic{}'s default background estimator (Section \ref{sec: background}) and source finder (Section \ref{sec: source identification}) routines, but used a custom \texttt{AperturePhotometer} instance (Section \ref{sec: photometry}) to define a circular aperture with a radius of 7 pixels. We also did not enable cosmic ray removal or dark current corrections. We then configured \texttt{Source-Extractor} to use \phoptic{}'s default parameter values where possible. Namely, we set the \texttt{DETECT\_MINAREA} parameter, which is equivalent to the \texttt{n\_pixels} parameter of \texttt{photutils.segmentation.SourceFinder} (Section \ref{sec: source identification}), to $128 / 2^2 = 32$ pixels, the \texttt{DETECT\_THRESH} and \texttt{ANALYSIS\_THRESH} parameters, which are equivalent to the \texttt{threshold} parameter of \phoptic{}'s \texttt{Reducer} class, to 5, the \texttt{PHOT\_APERTURES} parameter to 14 pixels (since \texttt{Source-Extractor} defines apertures in terms of diameter), and the \texttt{BACK\_SIZE} parameter, which is equivalent to the \texttt{box\_size} parameter of \texttt{photutils.background.Background2D} (Section \ref{sec: background}), to $1024 / 32 = 32$ pixels. We note, however, that there are some key differences between \phoptic{} and \texttt{Source-Extractor}. Most crucially, \phoptic{} and \texttt{Source-Extractor} use different sigma clipping routines to estimate the background, and both programs treat partially overlapping pixels differently when performing aperture photometry.

When estimating two-dimensional backgrounds, \texttt{Source-Extractor} (as of v2.28.2) iteratively clips background mesh boxes until all pixels are within $3 \sigma$ of the median value. In contrast, we recall that \phoptic{}'s default background estimator uses the same threshold, but only performs a maximum of 10 iterations. As such, \phoptic{}'s default background estimator will generally produce slightly different background estimates compared to \texttt{Source-Extractor}. We note that the \texttt{photutils.background} module, used by \phoptic{} to compute image backgrounds (Section \ref{sec: background}), can be configured to perform identically to \texttt{Source-Extractor}, but this is not how it is implemented in \phoptic{}.

When performing aperture photometry, \texttt{Source-Extractor} divides pixels into 5$\times$5 identical subpixels. A subpixel's flux is counted if its centre falls within the aperture. In contrast, the \texttt{photutils.aperture} module, which is used by \phoptic{} to perform aperture photometry, defaults to computing the fractional overlap of pixels and scaling their flux contributions accordingly. As such, \phoptic{} will generally produce slightly different flux measurements compared to \texttt{Source-Extractor}. We note that the \texttt{photutils.aperture} module can be configured to perform identically to \texttt{Source-Extractor}, but this is not how it is implemented in \phoptic{}.

\begin{figure}
    \centering
    \includegraphics[width=\columnwidth]{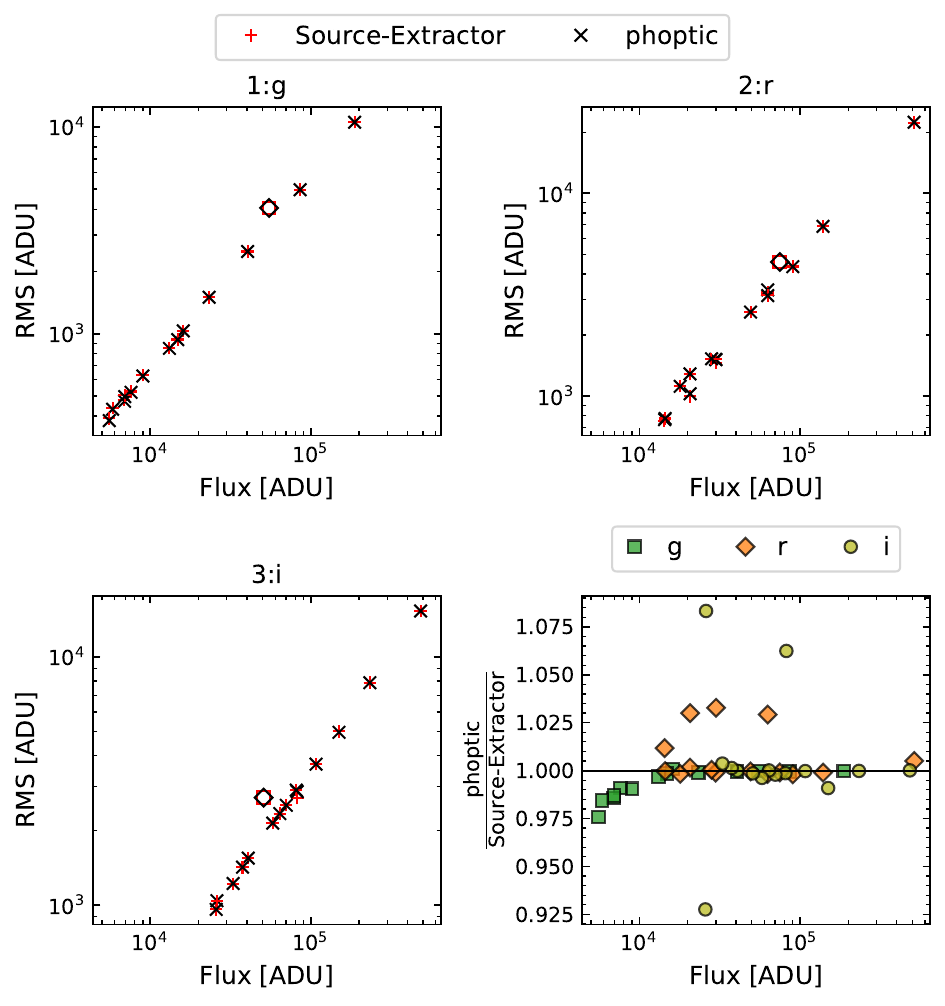}
    \caption{Top left, top right and bottom left: flux RMS as a function of flux for each camera from the observation represented in Figure \ref{fig: source catalogue}. ``x'' markers represent \phoptic{}'s measurements while ``+` markers represent \texttt{Source-Extractor}'s measurements. The variable source V709 Cas is identified with diamond and square markers for \phoptic{} and \texttt{Source-Extractor}, respectively. Bottom right: RMS ratios between \phoptic{} and \texttt{Source-Extractor} as functions of flux for each camera.}
    \label{fig: phoptic vs sextractor}
\end{figure}

In Figure \ref{fig: phoptic vs sextractor}, we show the ratio between \phoptic{} and \texttt{Source-Extractor}'s flux RMS values as a function of flux for the observation represented in Figure \ref{fig: source catalogue}. As can be seen from this figure, \phoptic{}'s results are very similar to those of \texttt{Source-Extractor}. Quantitatively, the median RMS ratios were 0.9977, 0.9998, and 0.9998 for the $g$-, $r$-, and $i$-bands, respectively, while we found extremes of 0.9757 ($g$-band), 1.0327 ($r$-band), and 1.0833 ($i$-band). On average, \phoptic{}'s results are therefore well within 1 per cent of those of \texttt{Source-Extractor} when both programs are configured similarly.

\section{Data analysis}\label{sec: data analysis}

\subsection{Computing relative light curves}\label{sec: differential photometry}

After reducing a series of science images, the resulting light curves will generally be dominated by atmospheric variability. As such, these raw light curves may have little scientific value on their own. As mentioned in Section \ref{sec: intro}, however, differential photometry can be used to largely mitigate atmospheric variability; \phoptic{} streamlines this process via the \texttt{DifferentialPhotometer} class. To compute a relative light curve, users must specify a camera--filter combination, and the corresponding target and comparison IDs.

\begin{figure*}
    \centering
    \includegraphics[width=\textwidth]{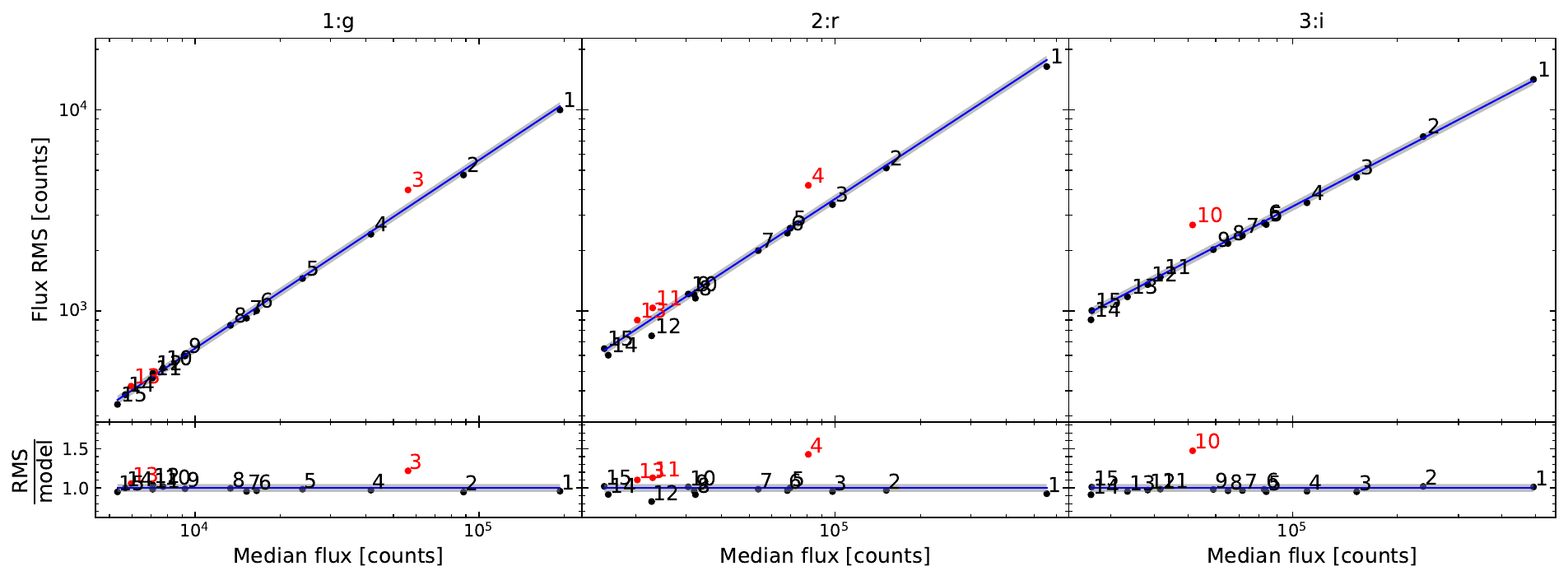}
    \caption{RMS--flux relations for the catalogued sources from Figure \ref{fig: source catalogue} as computed using aperture photometry. The solid lines show the power law fits to the data, while the shaded regions represent a $\pm 5$ per cent tolerance on the fits. Sources with RMS values more than 5 per cent above the power law fit are highlighted in red.}
    \label{fig: RMS vs flux example}
\end{figure*}

To aid in choosing suitable comparison sources when performing differential photometry, \phoptic{} generates RMS--flux plots for each catalogue after performing photometry. A power law is then fit to each catalogue's RMS--flux relation under the assumption that most sources in the field will be non-variable. Sources that have an unusually high RMS at a given flux are likely variable sources. In the RMS--flux plots, variable sources are highlighted in red. For example, Figure \ref{fig: RMS vs flux example} shows the RMS--flux plots corresponding to the source catalogues shown in Figure \ref{fig: source catalogue}. Any sources that are not flagged as variable are likely to be good comparison sources for performing differential photometry.

For convenience, users may specify \texttt{match\_other\_cameras=True} when computing the relative light curves. When \texttt{match\_other\_cameras=True}, \phoptic{} uses \texttt{astroalign} to match the target and comparison IDs from the input catalogue to any additional catalogues; thereby computing relative light curves for the target in multiple bands simultaneously. However, this may fail or misidentify sources, depending on the field, and so source matching should always be verified manually. After computing one or more relative light curves, an \texttt{phoptic.Analyzer} instance is returned.

\subsection{Quick-look analyses}\label{sec: analyzer}

\phoptic's \texttt{Analyzer} class contains convenience routines for handling and manipulating light curves, and performing a number of ``quick-look'' timing analyses. In terms of handling and manipulation, \texttt{Analyzer} instances can be combined to concatenate light curves from different nights, add light curves from new filters, and rebin light curves to a lower time resolution. Regarding quick-look analyses, \phoptic{} implements convenience routines for computing Lomb--Scargle periodograms (LSPs; \citealt{Lomb1976, Scargle1982}) and phase binning/folding light curves.

\section{Compatibility with other instruments}\label{sec: instruments}

\phoptic{} includes an \texttt{Instrument} data class that acts as an interface between \phoptic{} and the data products of a particular instrument. An \texttt{Instrument} instance defines the header keywords listed in Table \ref{tab: instrument keywords}, as well as the methods used to parse the values of these keywords into a standardised format.

\begin{table*}
    \centering
    \caption{\texttt{phoptic.Instrument}'s header keyword parameters.}
    \begin{tabularx}{\textwidth}{cccX}
        \hline
        Parameter & Default & Method & Description \\
        \hline
        \texttt{airmass\_kw} & AIRMASS & \texttt{get\_airmass()} & Reports the airmass of the image.\\
        \texttt{binning\_kw} & BINNING & \texttt{get\_binning()} & Reports the binning mode of the image (e.g., ``2$\times$2'').\\
        \texttt{camera\_kw} & INSTRUME & \texttt{get\_camera()} & Reports the camera/instrument used to take the image. This is used to apply bias and dark noise calibrations, and should not be ambiguous for multi-camera instruments.\\
        \texttt{dark\_curr\_kw} & DARKCURR & \texttt{get\_dark\_flux()} & Reports the dark current of the detector in electrons/pixel. This is multiplied by the exposure time to estimate the dark noise flux contribution.\\
        \texttt{dateobs\_kw} & DATEOBS & \texttt{get\_mjd()} & Reports the timestamp of the image. The \texttt{get\_mjd()} method is used to convert the image's timestamp into MJD format.\\
        \texttt{dec\_kw} & DEC & \texttt{get\_sky\_coord()} & Reports the declination angle of the image. This is used alongside the right ascension angle to apply a barycentric correction to the image's timestamp.\\
        \texttt{exptime\_kw} & EXPTIME & \texttt{get\_exptime()} & Reports the exposure time of the image in seconds.\\
        \texttt{filter\_kw} & FILTER & \texttt{get\_filter()} & Reports the filter used to take the image. This is used along with the camera to apply flat-field corrections and group images into catalogues.\\
        \texttt{gain\_kw} & GAIN & \texttt{get\_gain()} & Reports the gain of the detector in electrons/ADU.\\
        \texttt{ra\_kw} & RA & \texttt{get\_sky\_coord()} & Reports the right ascension angle of the image. This is used alongside the declination angle to apply a barycentric correction to the image's timestamp.\\
        \texttt{read\_noise\_kw} & RDNOISE & \texttt{get\_read\_noise()} & Reports the read noise of the detector in electrons/pixel.\\
        \hline
    \end{tabularx}
    \label{tab: instrument keywords}
    {NOTES:
    The ``Default'' column lists the default keywords assumed by \phoptic{}.
    The ``Method'' column lists the corresponding method used to parse the header keyword value into a standard format.
    The ``Description'' column describes what information the keyword should represent and how it is used by \phoptic{}.
    }
\end{table*}

In order to reduce data from another instrument using \phoptic{}, users will therefore need to subclass \texttt{Instrument}. To do this, \phoptic{} can be used to generate a JSON template; users may then edit the required keywords, and use the edited template to create a custom \texttt{Instrument} instance. Alternatively, users may write a custom \texttt{Instrument} subclass. Subclassing \texttt{Instrument} is much more flexible than instantiating an \texttt{Instrument} from a template file, and is particularly useful for instruments with ``non-standard'' headers. For example, the \texttt{get\_mjd()} method of the \texttt{Instrument} base class assumes timestamps are listed in FITS format (YYYY-MM-DDTHH:MM:SS.sss). If this is not the case for a particular instrument, a custom \texttt{get\_mjd()} method will need to be written to convert the listed timestamp into MJD format. Since the value returned by each header keyword is parsed using its respective method, there is no requirement for an instrument's image headers to include all the keywords listed in Table \ref{tab: instrument keywords}; instead, custom methods may be written for missing keywords to infer the required quantity from other keywords, or defined fixed values as required.

\section{Compatibility with other software}\label{sec: compatibility}

As detailed in Sections \ref{sec: data reduction} and \ref{sec: data analysis}, \phoptic{} makes use of a number of popular Python packages already used in astronomy. Notably, \texttt{astropy} and \texttt{photutils} are tightly-integrated into much of \phoptic{}'s core functionality. \phoptic{} also avoids creating binary files, instead saving files using generic/domain-specific filesystem formats such as (E)CSV, FITS, JSON, and TXT. It should therefore be straightforward for users to port \phoptic{} data products to their preferred analysis software.

\subsection{\texttt{Analyzer} as an interface to \texttt{stingray}}\label{sec: stingray}

To access more advanced timing analyses than those provided by the \texttt{Analyzer} class, light curves may be exported to \texttt{stingray} via the \texttt{export\_light\_curves\_to\_stingray()} method. \texttt{stingray} is a popular and feature-rich spectral timing Python package \citep[][]{stingray, stingray2} that implements advanced timing analysis routines, such as computing cross spectra, and includes a data modelling framework. However, since \texttt{stingray} was primarily written with X-ray data in mind, some routines may not be compatible with, or may not behave as expected when applied to, relative light curves (see Section \ref{sec: stingray limitations} for more details).

\subsubsection{Good Time Intervals}\label{sec: GTIs}

A number of \texttt{stingray}'s timing analysis routines require the specification of Good Time Intervals (GTIs). GTIs distinguish non-detections from observing gaps, and are extremely important for computing quantities such as averaged power spectra. To generate informative GTIs, \phoptic{} uses a heuristic to filter out gaps. A gap is identified if the time between consecutive data points in a light curve is greater than $1.5 \delta t$, where $\delta t$ is the median time between consecutive data points in the light curve. The start of a gap then becomes the end of one GTI, and the end of the gap becomes the start of the next GTI. The default GTIs inferred by \phoptic{} therefore do not distinguish non-detections from observing gaps, which may not be desired behaviour in all cases.

\subsubsection{Limitations}\label{sec: stingray limitations}

As mentioned above, \texttt{stingray} was primarily developed for use with X-ray data. \texttt{stingray} therefore makes several assumptions when creating light curves, most notably that fluxes follow a Poisson error distribution. We note that, at the time of writing, \texttt{stingray} includes partial support for other error distributions with the following warning: ``Stingray only uses poisson err\_dist at the moment. All analyses in the light curve will assume Poisson errors. Sorry for the inconvenience.'' This is problematic for relative light curves since they follow a Gaussian error distribution.

A consequence of the above is that some of \texttt{stingray}'s routines will not work as expected when applied to relative light curves, if at all. For example, it is possible to compute the cross-spectrum between two relative light curves in \texttt{stingray}. However, when attempting to compute the coherence or lags from the cross-spectrum, unphysical or \texttt{NaN} values may be produced. This is because the coherence, which is used to derive lags, depends on the Poisson noise levels of the underlying power spectra \citep[see, for example,][]{Vaughan1997}. If the light curve is relative, such that it has non-integer flux values, a Poisson error distribution is ill-defined, the Poisson noise level cannot be estimated reliably, and unphysical coherence values are produced.

For some of \texttt{stingray}'s routines, non-Poisson error distributions may be used without issue, provided some care is taken in regards to units. For example, when computing power spectra, certain normalisations (such as Leahy normalisation; \citealt{Leahy1983}) will not behave as expected when applied to relative light curves, while scale-invariant normalisations (such as fractional RMS normalisation; \citealt{Belloni1990}) can be used without issue.

\section{Examples}\label{sec: example}

In this section, we demonstrate how to use \phoptic{} to reduce observations from OPTICAM (Section \ref{sec: OPTICAM examples}), ULTRACAM (Section \ref{sec: ULTRACAM example}), HiPERCAM (Section \ref{sec: HiPERCAM example}), and MEXMAN (Section \ref{sec: MEXMAN example}). We also perform some relevant ``quick-look'' analyses on the reduced data products, but keep our discussion largely phenomenological.

\subsection{OPTICAM}\label{sec: OPTICAM examples}

Since \phoptic{} was designed primarily for OPTICAM, these observations are the most straightforward to reduce. To demonstrate the capabilities of both \phoptic{} and OPTICAM, we provide two examples: in the first example, we reduce observations of an eclipsing source in a sparse field; in the second example, we reduce observations of a somewhat crowded field containing a reasonably bright source with strongly wavelength-dependent emission.

\subsubsection{DO Leo}\label{sec: DO Leo example}

For our first OPTICAM example, we reproduce the light curves shown in \cite{Castro2024b} of the eclipsing nova-like cataclysmic variable DO Leo. DO Leo was observed on May 16th, 2022, using the $g$, $r$, and $i$ filters in the 2$\times$2 binning mode; since DO Leo is a somewhat faint source (16.9 $G$ mag; \citealt{GaiaEDR3}), it was observed using an exposure time of 30~s. For each camera, 252 images were taken.

For the reduction parameters, we applied the 3$\times$3 median filter recommended by \cite{Paez2026}, and enabled cosmic ray removal. Additionally, we applied dark noise and flat-field corrections using the respective routines (Section \ref{sec: corrections}). Since these observations were taken in 2022, the image headers do not contain coordinate information; as such, we disabled barycentric corrections. We also note coordinate information is required to estimate the airmass, which is used to estimate the scintillation noise (Section \ref{sec: error propagation}). For this observation, the scintillation noise is therefore assumed to be zero; given the long exposure time, in addition to DO Leo being a somewhat faint source, this is likely a reasonable approximation. All other reduction parameters were left to their default values (Table \ref{tab: reducer params}).

\begin{figure*}
    \centering
    \includegraphics[width=\textwidth]{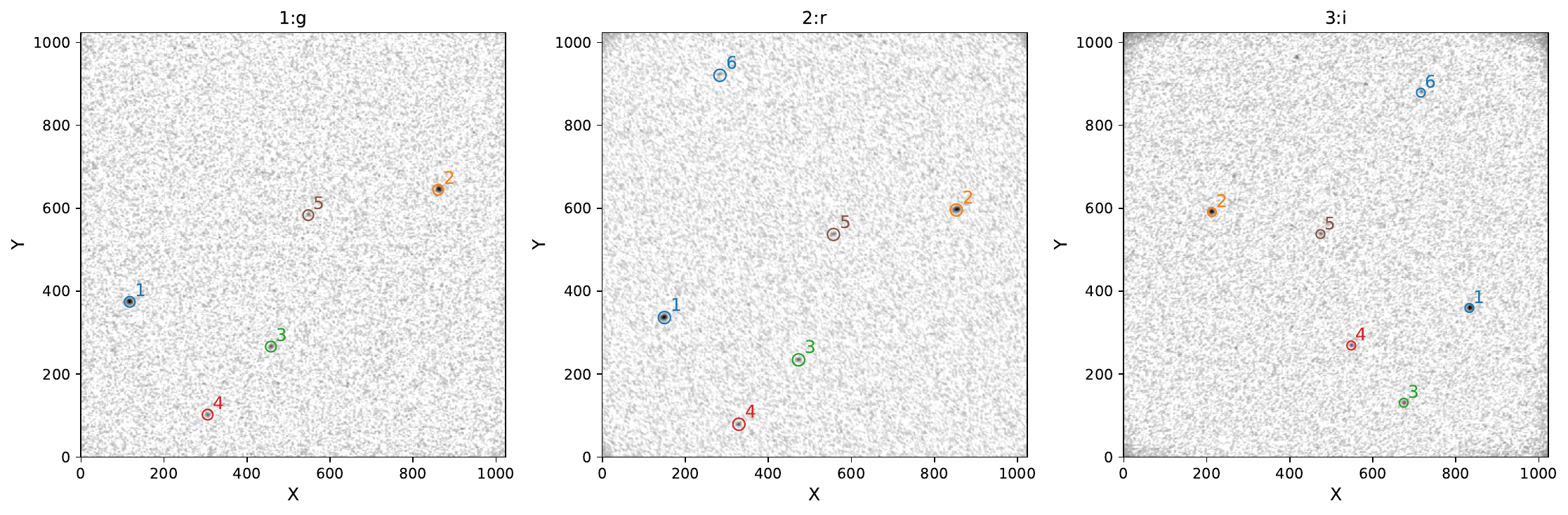}
    \caption{Source catalogues for DO Leo observations. DO Leo corresponds to $g$-band Source 3, $r$-band Source 3, and $i$-band Source 4. As with most observations from 2022, Camera 3's images are flipped due to technical issues.}
    \label{fig: DO Leo catalogues}
\end{figure*}

To construct source catalogues, affine transformations cannot be used as there are too few sufficiently bright sources. Instead, we therefore used \texttt{transform\_type="translation"} and \texttt{n\_alignment\_sources=1} (Section \ref{sec: source catalogues}). We also imposed a translation limit of 64 pixels since the final exposure of this observation is severely misaligned. The resulting source catalogues are presented in Figure \ref{fig: DO Leo catalogues}, with DO Leo corresponding to Source 3 in the $g$ and $r$ catalogues, and Source 4 in the $i$ catalogue.

Since DO Leo is an eclipsing source, it may exhibit large drops in S/N and be undetectable in some images. To prevent gaps in the resulting light curves, we therefore performed forced photometry using the default aperture and optimal photometers (Section \ref{sec: photometry}). For both types of photometry, we estimated the background locally using \phoptic's default annulus (Section \ref{sec: local background}). The full reduction script is provided in Listing \ref{listing: DO Leo}.

\begin{lstlisting}[caption={Data reduction script for an OPTICAM observation of DO Leo.},language=Python,label={listing: DO Leo}]
import os

os.environ['OMP_NUM_THREADS'] = '1'

from functools import partial

from scipy.ndimage import median_filter

import phoptic


def main():
    image_filter = partial(median_filter, size=3, mode='mirror')  # 3x3 median filter recommended by Praez+2026
    
    dark_corr = phoptic.DarkNoiseCorrector()
    
    flat_corr = phoptic.FlatFieldCorrector(
        out_directory=os.path.join(os.path.dirname(__file__), 'Reduced/20220516/Calibrators'),
        data_directory='/mnt/SATA1/Data/Opticam/Raw/Flats/20220516/2x2/0_05s',
        dark_corrector=dark_corr,
        image_filter=image_filter,
        )
    
    reducer = phoptic.Reducer(
        out_directory=os.path.join(os.path.dirname(__file__), 'Reduced/20220516'),
        data_directory='/mnt/SATA1/Data/Opticam/Raw/DO_Leo/20220516',
        dark_corrector=dark_corr,
        flat_corrector=flat_corr,
        remove_cosmic_rays=True,
        show_plots=False,
        barycenter=False,
        image_filter=image_filter,
        )
    
    reducer.create_catalogs(
        transform_type='translation',
        n_alignment_sources=1,
        translation_limit=64,
    )
    
    annulus = phoptic.DefaultLocalBackground()
    photometer = phoptic.OptimalPhotometer(
        forced=True,
        local_background_estimator=annulus,
    )
    reducer.photometry(photometer)
    
    photometer = phoptic.AperturePhotometer(
        forced=True,
        local_background_estimator=annulus,
    )
    reducer.photometry(photometer)


if __name__ == '__main__':
    main()
\end{lstlisting}

\begin{figure}
    \centering
    \includegraphics[width=\columnwidth]{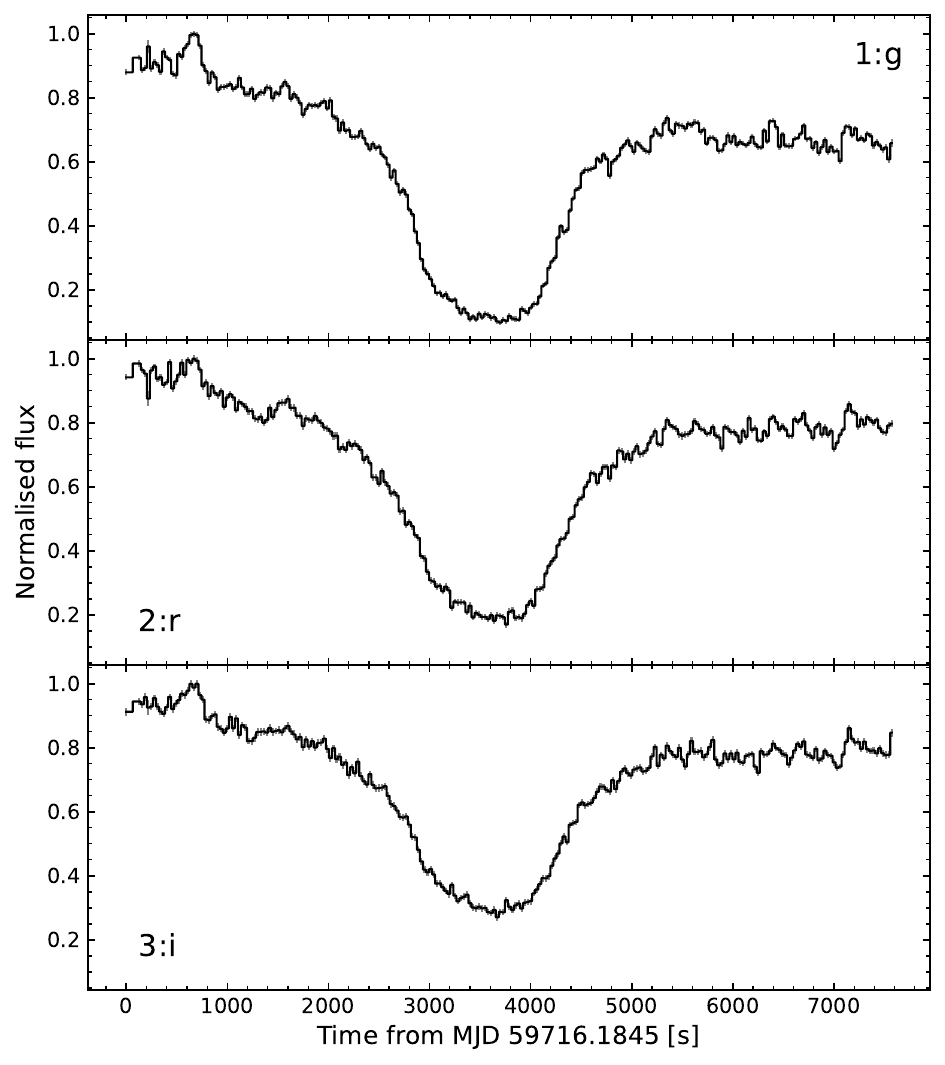}
    \caption{OPTICAM light curves of DO Leo from May 16th 2022.}
    \label{fig: DO Leo eclipse}
\end{figure}

To compute relative light curves, we used Sources 1 and 2 (which correspond to the same sources in all three catalogues) as comparison sources. The resulting light curves, computed using optimal photometry, are presented in Figure \ref{fig: DO Leo eclipse}.

\subsubsection{V709 Cas}\label{sec: V709 Cas example}

For our second OPTICAM example, we reduce an observation of the intermediate polar V709 Cas taken on November 4th, 2024. V709 Cas is a reasonably bright source ($G$ mag 14.4; \citealt{GaiaEDR3}) in a somewhat crowded field. For this particular observation, V709 Cas was observed using the $g$, $r$, and $i$ filters with 3~s exposures in the 2$\times$2 binning mode. For each camera, 3501 images were taken.

To reduce these data, we performed dark noise corrections (Section \ref{sec: dark noise corrections}) and set \texttt{remove\_cosmic\_rays=True}. We omit flat-field corrections since we found that they made a negligible difference to the final light curves in this case. All other reduction parameters were left to their default values (Table \ref{tab: reducer params}).

We then constructed source catalogues using \phoptic{}'s default values. The resulting source catalogues are shown in Figure \ref{fig: source catalogue}, with V709 Cas corresponding to $g$-band Source 3, $r$-band Source 4, and $i$-band Source 10. We recall that the noise characterisation plots for this observation are shown in Figure \ref{fig: noise characterisation}, where it can be seen that V709 Cas is shot-noise-dominated in all three bands.

\begin{figure}
    \centering
    \includegraphics[width=\columnwidth]{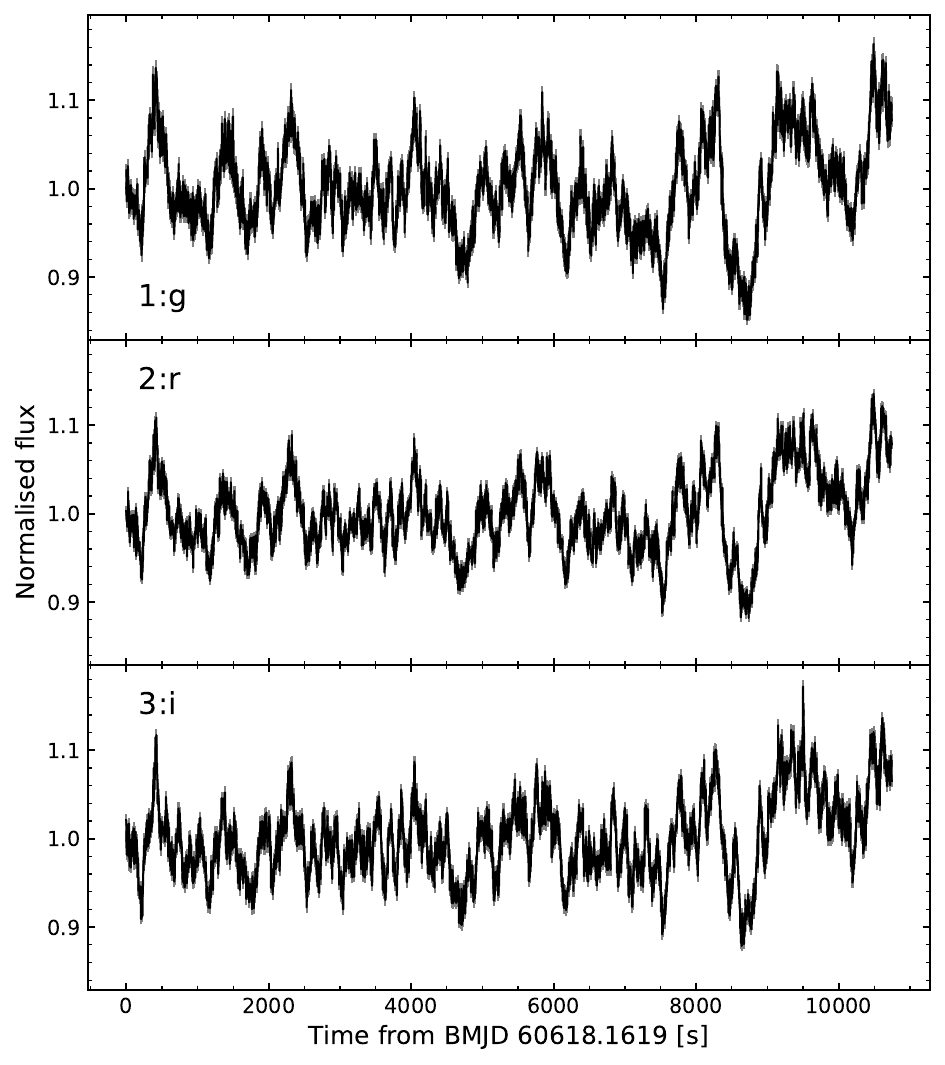}
    \caption{OPTICAM light curves of V709 Cas from November 4th, 2024, computed using optimal photometry.}
    \label{fig: V709 Cas relative light curves}
\end{figure}

To perform photometry, we used the default aperture and optimal photometers (Section \ref{sec: photometry}). From Figure \ref{fig: RMS vs flux example} it can be seen that V709 Cas has been flagged as variable in all bands, but appears most variable in the $i$-band. Close to V709 Cas in the field is a source of similar brightness ($g$-band Source 4; $r$-band Source 3; $i$-band Source 6) that does not appear to show any intrinsic variability. As such, we computed relative light curves for V709 Cas using this nearby source for comparison. The resulting light curves, as computed using optimal photometry, are presented in Figure \ref{fig: V709 Cas relative light curves}. The full reduction script is provided in Listing \ref{listing: V709 Cas}.

\begin{lstlisting}[caption={Data reduction script for an OPTICAM observation of V709 Cas.},language=Python,label={listing: V709 Cas}]
import os

os.environ['OMP_NUM_THREADS'] = '1'

import phoptic


def main():
    dark_corr = phoptic.DarkNoiseCorrector()
    
    reducer = phoptic.Reducer(
        out_directory=os.path.join(os.path.dirname(__file__), 'Reduced/20241104'),
        data_directory='/mnt/SATA1/Data/Opticam/Raw/V709_Cas/20241104',
        dark_corrector=dark_corr,
        remove_cosmic_rays=True,
        show_plots=False,
        )
    
    reducer.create_catalogs()
    
    photometer = phoptic.OptimalPhotometer()
    reducer.photometry(photometer)
    
    photometer = phoptic.AperturePhotometer()
    reducer.photometry(photometer)


if __name__ == '__main__':
    main()
\end{lstlisting}

\begin{figure}
    \centering
    \includegraphics[width=\columnwidth]{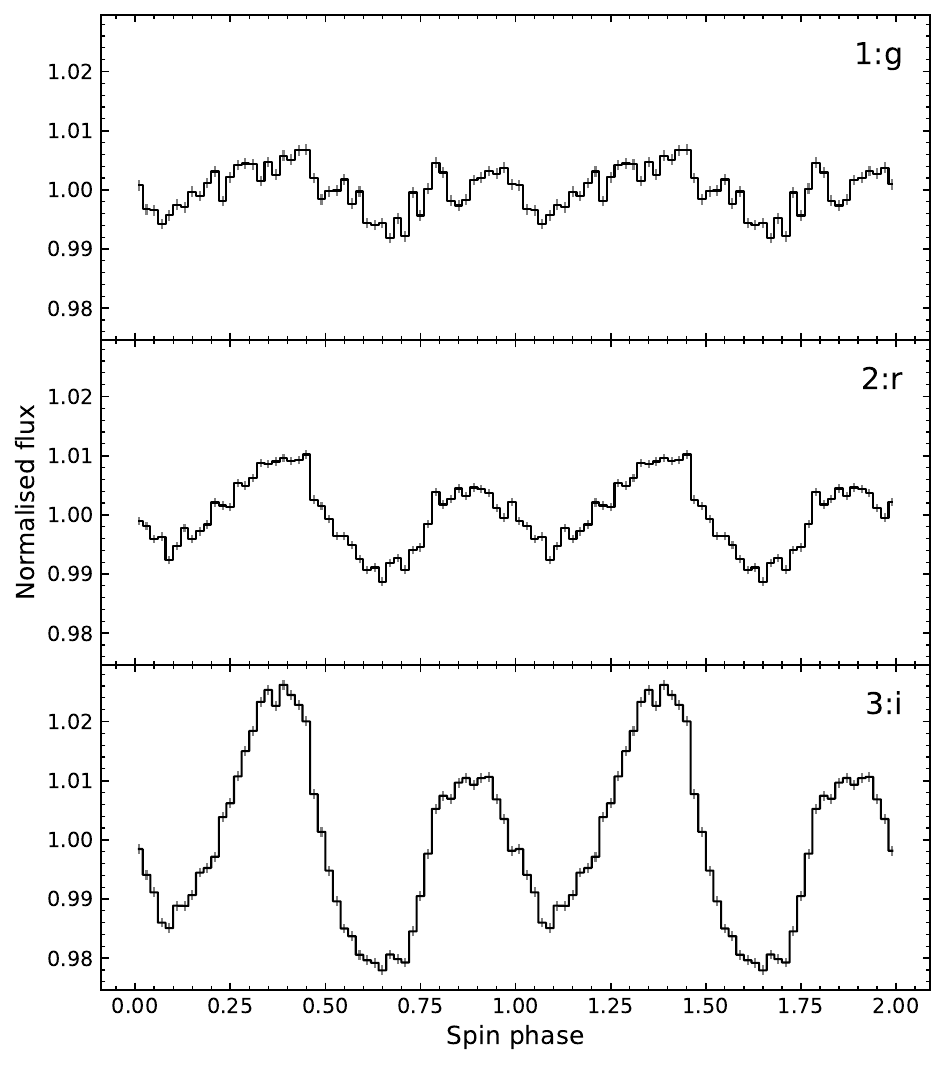}
    \caption{Light curves from Figure \ref{fig: V709 Cas relative light curves} folded on V709 Cas's 312.75~s spin period and binned in phase. Two periods are shown for clarity.}
    \label{fig: V709 Cas phase bin}
\end{figure}

The spin period of V709 Cas is known to be 312.75~s \citep[e.g.,][]{Rao2025}. Using this period, we fold and bin the light curves from Figure \ref{fig: V709 Cas relative light curves} to visualise the pulse profile at different wavelengths; we arbitrarily set the time of zero phase to the time of the first exposure, BMJD 60618.1619. The resulting pulse profiles are presented in Figure \ref{fig: V709 Cas phase bin}.

As can be seen from Figure \ref{fig: V709 Cas phase bin}, the $i$-band spin pulsations are considerably larger amplitude than those of the $g$- and $r$-bands. \cite{Irving2026} further showed that the $z$-band spin pulsations are even larger amplitude than those of the $i$-band, which they interpreted as a signature of cyclotron emission.

\subsection{ULTRACAM}\label{sec: ULTRACAM example}

An instructive test for \phoptic{} is to see how it performs when reducing data from ULTRACAM \citep[][]{Dhillon2007}, a multi-camera instrument that has been used extensively to study a range of timing phenomena. For example, ULTRACAM has been used to observe low mass X-ray binaries \citep[e.g.,][]{Durant2008}, accreting white dwarfs \citep[e.g.,][]{Copperwheat2009}, exoplanets transits \citep[e.g.,][]{Frohring2013}, and  M dwarf flares \citep[e.g.,][]{Kowalski2016}. ULTRACAM is therefore a good benchmark instrument for a photometry pipeline.

OPTICAM is similar to ULTRACAM: both instruments are comprised of three cameras and are equipped with a set of SDSS $ugriz$ filters. Unlike OPTICAM, however, ULTRACAM uses traditional CCDs, \textbf{and is now permanently mounted on the New Technology Telescope in Chile}. Moreover, ULTRACAM implements three different observing modes: full-frame, windowed, and drift. Full-frame mode operates similarly to OPTICAM, and reads the entire imaging area each cycle. Windowed mode allows the observer to define regions of interest (i.e., ``windows'') on each CCD, reducing the readout area and improving frame-rates. Drift mode is similar to windowed mode, but is less flexible in order to achieve even higher frame-rates. ULTRACAM also produces custom-format FITS cubes, which must first be converted to standard-format FITS files for compatibility with programs like \phoptic{} and \texttt{Source-Extractor}.

In this section, we use \phoptic{} to reduce an ULTRACAM observation of the cataclysmic variable LU Cam made on January 10th, 2012 \citep[][]{Scaringi2013}. \textbf{When this observation was taken, ULTRACAM was mounted on the 4.2~m William Herschel Telescope in La Palma.} For this observation, ULTRACAM was used in windowed mode with two windows, the binning factor was set to 1, and the readout mode was set to slow. Additionally, the $u$-, $g$-, and $r$-band filters were used. For the $g$- and $r$-bands, a total of 18,502 exposures were taken using an exposure time of 0.5~s; to improve the S/N in the $u$-band, the CCD was read out every other exposure, thereby doubling the exposure time and halving the number of exposures.

Before we could reduce these data using \phoptic{}, we first converted them to standard-format FITS files using the \texttt{hipercam} software\footnote{\url{https://github.com/HiPERCAM/hipercam}}, which includes a \texttt{ucam} module for ULTRACAM data. Since \phoptic{} does not natively support windowed observing modes, we ``stitched'' the two windows together for each exposure of each CCD to form a single image. Stitching windows together in this way results in a discontinuity in the middle column of the resulting image, which may skew global background estimation. For sources away from this discontinuity, however, the background can be estimated locally without issue.

To enable \phoptic{} to interface with the standard-format ULTRACAM images, we then defined a custom \texttt{Instrument} subclass. To estimate the scintillation noise, we use equation \ref{eq: scintillation noise} with the median La Palma correction coefficient of 1.30 found by \cite{Osborn2015}. This \texttt{ULTRACAM} \texttt{Instrument} is provided in Listing \ref{listing: ULTRACAM instrument}.

\begin{lstlisting}[caption={\texttt{Instrument} subclass for the ULTRACAM instrument.},language=Python,label={listing: ULTRACAM instrument}]
from astropy.coordinates import AltAz, EarthLocation
from astropy.time import Time
from astropy import units as u
import numpy as np


from phoptic import Instrument


class ULTRACAM(Instrument):
    
    def __init__(
        self,
        readout_mode,
        ):
        
        super().__init__(
            diameter=4.2 * u.m,
            location=EarthLocation.from_geodetic(
                lon=-17.882 * u.deg,
                lat=28.761 * u.deg,
                height=2396 * u.m,
                ),
            pixel_scales={
                'Red CCD': 0.3,
                'Green CCD': 0.3,
                'Blue CCD': 0.3,
                },
            dateobs_kw='MJDINT',
            camera_kw='CCDNAME',
            binning_kw='CCDSUM',
            )
        
        assert readout_mode in ['fast', 'slow'], 'readout_mode must be one of: "fast", "slow"'
        self.readout_mode = readout_mode


    def get_mjd(
        self,
        file = None,
        header = None,
        ):
        
        if header is None:
            header = file.get_header()
        
        return float(header['MJDINT']) + float(header['MJDFRAC'])


    def get_read_noise(
        self,
        file = None,
        header = None,
        ):
        
        if self.readout_mode == 'fast':
            return 5.0
        else:
            return 3.5


    def get_airmass(
        self,
        file = None,
        header = None,
        ):
        
        if header is None:
            header = file.get_header()
        
        sky_coord = self.get_sky_coord(file=file, header=header)
        obstime = Time(self.get_mjd(file=file, header=header), format='mjd')
        
        altaz = sky_coord.transform_to(AltAz(obstime=obstime, location=self.location))
        
        return 1 / np.sin(altaz.alt.rad)


    def get_relative_scintillation_noise(
        self,
        file = None,
        header = None,
        C = 1.30,
        H = 8000 * u.m,
        ):
        
        return super().get_relative_scintillation_noise(file=file, header=header, C=C, H=H)
\end{lstlisting}

Next, we defined bias and flat-field correctors using the associated calibration frames (Section \ref{sec: corrections}), but neglected dark noise corrections (which is a negligible noise source for ULTRACAM; \citealt{Dhillon2007}). To estimate the global image backgrounds, we used \phoptic{}'s default background estimator with the \texttt{box\_size} parameter set to 20 pixels (Section \ref{sec: background}). For source identification, we used \phoptic{}'s default routine with \texttt{n\_pixels = 16} (Section \ref{sec: source identification}). When instantiating the \texttt{Reducer}, we also enabled cosmic ray removal before computing source catalogues using the default parameters (Section \ref{sec: source catalogues}).

After constructing source catalogues, we performed aperture photometry, using the default aperture size (Section \ref{sec: photometry}), with \phoptic{}'s default background annulus (Section \ref{sec: local background}). The full reduction script is provided in Listing \ref{listing: ULTRACAM example}.

\begin{lstlisting}[caption={Data reduction script for the ULTRACAM observation described in Section \ref{sec: ULTRACAM example}.},language=Python,label={listing: ULTRACAM example}]
import os
from pathlib import Path

os.environ['OMP_NUM_THREADS'] = '1'

import phoptic

from instrument import ULTRACAM


DATA_DIR = Path('/mnt/SATA2/Data/ULTRACAM/LU_Cam')
OUT_DIR = Path('Reduced')


def main():
    instrument = ULTRACAM(readout_mode='slow')
    
    bias_corr = phoptic.BiasCorrector(
        out_directory=OUT_DIR / 'correctors' / 'bias',
        data_directory=DATA_DIR / 'Bias',
        instrument=instrument,
        )
    
    flat_corr = phoptic.FlatFieldCorrector(
        out_directory=OUT_DIR / 'correctors' / 'flat',
        data_directory=DATA_DIR / 'Flats',
        bias_corrector=bias_corr,
        instrument=instrument,
        )
    
    background = phoptic.DefaultBackground(20)
    finder = phoptic.DefaultFinder(16, 16)
    
    reducer = phoptic.Reducer(
        out_directory=OUT_DIR,
        data_directory=DATA_DIR / 'Object',
        remove_cosmic_rays=True,
        show_plots=False,
        bias_corrector=bias_corr,
        flat_corrector=flat_corr,
        instrument=instrument,
        background=background,
        finder=finder,
        )
    
    reducer.create_catalogs()
    
    annulus = phoptic.DefaultLocalBackground()
    photometer = phoptic.AperturePhotometer(
        local_background_estimator=annulus,
        )
    reducer.photometry(photometer)


if __name__ == '__main__':
    main()
\end{lstlisting}

\begin{figure*}
    \centering
    \includegraphics[width=\textwidth]{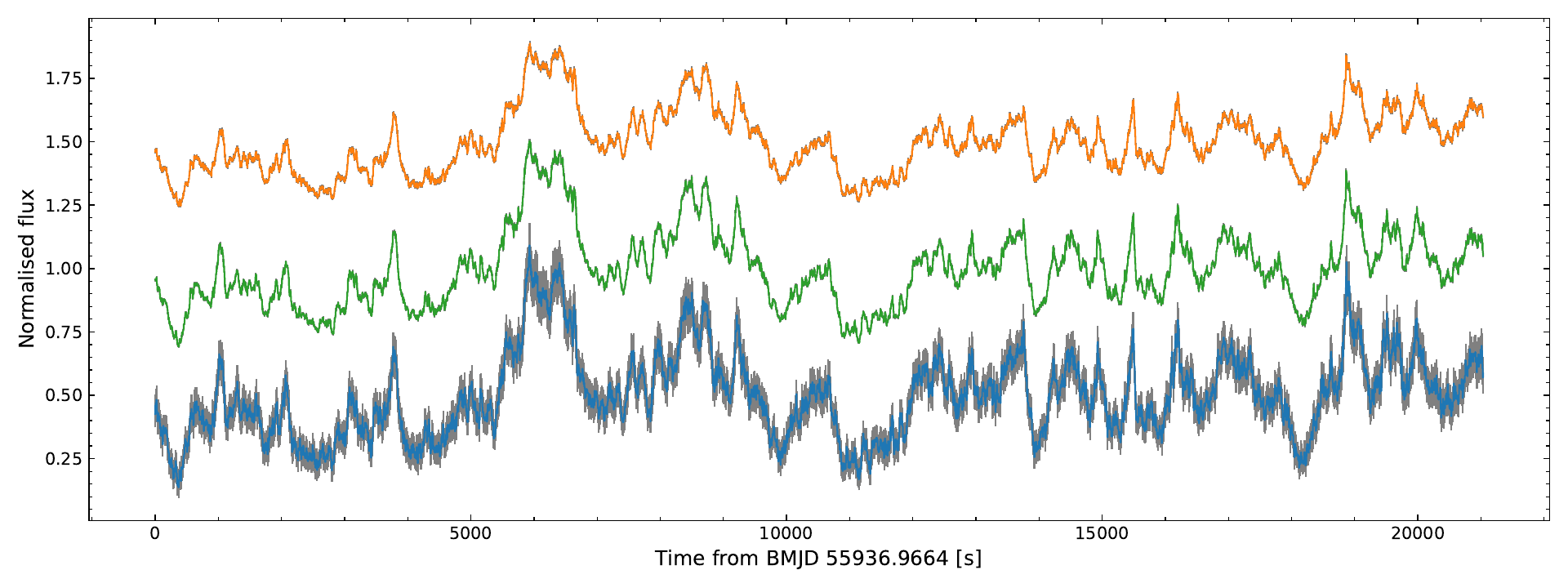}
    \caption{ULTRACAM light curves of LU Cam produced by \phoptic{}. All three light curves have been normalised to a mean flux of 1. The top light curve represents the $r$-band (with a +0.5 flux offset); the middle light curve represents the $g$-band; the bottom light curve represents the $u$-band (with a -0.5 flux offset).}
    \label{fig: LU Cam ULTRACAM light curves}
\end{figure*}

We then computed relative light curves using a non-varying comparison source at RA 05:58:31.6 DEC +67:52:42.0. The resulting light curves, binned to a time resolution of 2~s, are presented in Figure \ref{fig: LU Cam ULTRACAM light curves}.

By comparing Figure \ref{fig: LU Cam ULTRACAM light curves} to Figure 1 of \cite{Scaringi2013}, which was produced using the dedicated ULTRACAM pipeline \citep{Dhillon2007}, it can be seen that both sets of light curves are similar. It is therefore clear that \phoptic{} can successfully reduce ULTRACAM data - though only once they have been converted to standard-format FITS images.

\subsection{HiPERCAM}\label{sec: HiPERCAM example}

Designed to improve upon ULTRACAM, HiPERCAM is a five-camera instrument mounted on the 10.4~m Gran Telescopio Canarias (GTC) in La Palma, and is equipped with a set of custom $u_sg_sr_si_sz_s$ filters \citep{Dhillon2021}. HiPERCAM can therefore be used to demonstrate \phoptic{}'s ability to scale to a higher numbers of cameras. HiPERCAM's $u_sg_sr_si_sz_s$ filters are similar to the standard $ugriz$ filters used by OPTICAM and ULTRACAM, but with higher throughput ($\sim$~5--10 per cent for the $g_zr_si_sz_z$ filters, and 41 per cent for the $u_s$ filter). Similarly to ULTRACAM (Section \ref{sec: ULTRACAM example}), HiPERCAM also implements full-frame, windowed, and drift readout modes, and produces custom-format FITS cubes.

In this section, we use \phoptic{} to reduce a HiPERCAM observation of PSR J2051-0827 made on August 6th, 2021 \citep{Dhillon2022}. PSR J2051-0827 is an eclipsing millisecond pulsar with a 22.3 $R$ mag companion and a 2.4~hr orbital period \citep[e.g.,][]{Stappers1996}. For this observation, HiPERCAM was used in full-frame mode with a binning factor of 2. Additionally, the exposure time was set to 30.8~s (slow readout mode) and a total of 300 exposures were taken for each filter, with the exception of the $u_s$-band. To improve the S/N in the $u_s$-band, the exposure time was doubled, thereby halving the number of exposures.

The first step in our reduction was to use the \texttt{hipercam} software\footnote{\url{https://github.com/HiPERCAM/hipercam}} to convert the custom-format FITS file to a standard format. To then enable \phoptic{} to interface with these data, we defined a custom \texttt{Instrument} subclass. We estimate the scintillation noise in a similar way as for ULTRACAM (Section \ref{sec: ULTRACAM example}), since both instruments are at the same observatory.\footnote{\textbf{Note: ULTRACAM was in La Palma when the data reduced in Section \ref{sec: ULTRACAM example} were taken.}} This \texttt{HiPERCAM} \texttt{Instrument} is provided in Listing \ref{listing: HiPERCAM instrument}.

\begin{lstlisting}[caption={\texttt{Instrument} subclass for the HiPERCAM instrument.},language=Python,label={listing: HiPERCAM instrument}]
from astropy.coordinates import AltAz, EarthLocation
from astropy.time import Time
from astropy import units as u
import numpy as np


from phoptic import Instrument


class HiPERCAM(Instrument):
    
    def __init__(
        self,
        readout_mode,
        ):
        
        super().__init__(
            diameter=10.4 * u.m,
            location=EarthLocation.from_geodetic(
                lon=-17.882 * u.deg,
                lat=28.761 * u.deg,
                height=2396 * u.m,
                ),
            pixel_scales={
                '1': 0.081,
                '2': 0.081,
                '3': 0.081,
                '4': 0.081,
                '5': 0.081,
                },
            dateobs_kw='MJDINT',
            camera_kw='CCD',
            binning_kw='CCDSUM',
            )
        
        assert readout_mode in ['fast', 'slow'], 'readout_mode must be one of: "fast", "slow"'
        self.readout_mode = readout_mode


    def get_mjd(
        self,
        file = None,
        header = None,
        ):
        
        if header is None:
            header = file.get_header()
        
        return float(header['MJDINT']) + float(header['MJDFRAC'])


    def get_read_noise(
        self,
        file = None,
        header = None,
        ):
        
        if self.readout_mode == 'fast':
            return 5.4
        else:
            return 4.5


    def get_airmass(
        self,
        file = None,
        header = None,
        ):
        
        if header is None:
            header = file.get_header()
        
        sky_coord = self.get_sky_coord(file=file, header=header)
        obstime = Time(self.get_mjd(file=file, header=header), format='mjd')
        
        altaz = sky_coord.transform_to(AltAz(obstime=obstime, location=self.location))
        
        return 1 / np.sin(altaz.alt.rad)


    def get_relative_scintillation_noise(
        self,
        file = None,
        header = None,
        C = 1.30,
        H = 8000 * u.m,
        ):
        
        return super().get_relative_scintillation_noise(file=file, header=header, C=C, H=H)
\end{lstlisting}

With the HiPERCAM data in standard format and the \texttt{HiPERCAM} \texttt{Instrument} defined, we then proceeded with reduction. We defined bias and flat-field correctors (Section \ref{sec: corrections}), but neglect dark noise corrections due to HiPERCAM's detectors having a negligible dark current \citep{Dhillon2021}. To measure image backgrounds, we used \phoptic{}'s default background estimator (Section \ref{sec: background}) with its default parameter values. For source identification, we used \phoptic{}'s default routine with \texttt{n\_pixels = 32} (Section \ref{sec: source identification}). We also enabled cosmic ray removal.

After computing source catalogues for each CCD, we found that PSR J2051-0827 had not been catalogued. However, we recall that the companion to PSR J2051-0827 is very faint ($R$ 22.3; \citealt{Stappers1996}), and so this is not surprising. We therefore used \phoptic{}'s \texttt{pick\_sources()} routine to manually identify PSR J2051-0827 in each catalogue before performing forced aperture photometry (using the default aperture size; Section \ref{sec: photometry}). The full reduction script is provided in Listing \ref{listing: HiPERCAM example}.

\begin{lstlisting}[caption={Data reduction script for the HiPERCAM observation described in Section \ref{sec: HiPERCAM example}.},language=Python,label={listing: HiPERCAM example}]
import os
from pathlib import Path

os.environ['OMP_NUM_THREADS'] = '1'

import phoptic

from instrument import HiPERCAM


DATA_DIR = Path('Data/PSR_J2051')
OUT_DIR = Path('Reduced')


def main():
    instrument = HiPERCAM(readout_mode='slow')
    
    bias_corr = phoptic.BiasCorrector(
        out_directory=OUT_DIR / 'correctors' / 'bias',
        data_directory=DATA_DIR / 'bias',
        instrument=instrument,
        )
    
    flat_bias_corr = phoptic.BiasCorrector(
        out_directory=OUT_DIR / 'correctors' / 'flat_bias',
        data_directory=DATA_DIR / 'flat_bias',
        instrument=instrument,
        rebin_factor=2,
        )
    
    flat_corr = phoptic.FlatFieldCorrector(
        out_directory=OUT_DIR / 'correctors' / 'flat',
        data_directory=DATA_DIR / 'flat',
        bias_corrector=flat_bias_corr,
        instrument=instrument,
        rebin_factor=2,
        )
    
    finder = phoptic.DefaultFinder(32, 16)
    
    reducer = phoptic.Reducer(
        out_directory=OUT_DIR,
        data_directory=DATA_DIR / 'object',
        remove_cosmic_rays=True,
        show_plots=False,
        bias_corrector=bias_corr,
        flat_corrector=flat_corr,
        instrument=instrument,
        finder=finder,
        )
    
    reducer.create_catalogs()
    
    reducer.pick_sources()
    
    annulus = phoptic.DefaultLocalBackground()
    photometer = phoptic.AperturePhotometer(
        local_background_estimator=annulus,
        forced=True,
        )
    reducer.photometry(photometer)


if __name__ == '__main__':
    main()
\end{lstlisting}

We then computed relative light curves in each band using a nearby source at approximately RA 20:51:6.5 DEC -08:27:30.5, folded the resulting light curves on the 2.4~hr orbital period, and binned the folded light curves in phase. The time of zero phase was arbitrarily set to BMJD 59099.

\begin{figure}
    \centering
    \includegraphics[width=\columnwidth]{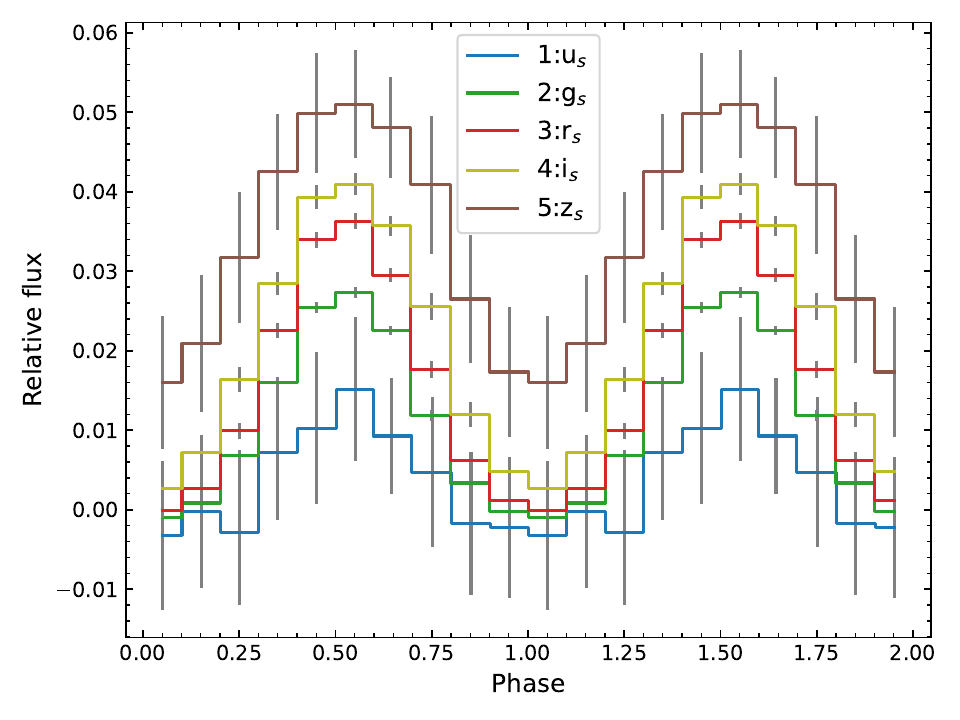}
    \caption{HiPERCAM light curves of PSR J2051-0827 folded on the 2.4~hr orbital period and binned in phase. From top to bottom, the filters are: $z_s$ (brown), $i_s$ (yellow), $r_s$ (red), $g_s$ (green), and $u_s$ (blue). The time of zero phase was set to BMJD 59099 and two periods are shown for clarity.}
    \label{fig: PSR J2051 light curves}
\end{figure}

As can be seen from Figure \ref{fig: PSR J2051 light curves}, the resulting light curves show clear modulations on the expected 2.4~hr orbital period, consistent with those shown in Figure 3 of \cite{Dhillon2022}. We note that even with the exposure time doubled, PSR J2051-0827 is still heavily sky-limited in the $u_s$-band, resulting in large error bars and negative fluxes at pulse minimum. Conversely, the error bars for the $z_s$ band are artificially increased due to fringing. Fringing can be corrected for in a similar way to flat-fielding (Section \ref{sec: flat corrections}), but \phoptic{} currently does not support applying additional corrections natively.

\subsection{MEXMAN}\label{sec: MEXMAN example}

Another instructive test for \phoptic{} is to see how it performs for single-camera instruments with multiple filters. In this section, we therefore use \phoptic{} to reduce an observation made on February 3rd, 2026, using the MEXMAN instrument mounted on the 0.84~m telescope at the OAN-SPM. Unlike OPTICAM, MEXMAN uses a single 2064$\times$2048 CCD and is equipped with a $UBVRI$ Bessel filter wheel. For this particular observation, MEXMAN was used to observe the eclipsing binary [GGM2006] 2532136 \citep[][]{Gettel2006} at RA 10:04:14.4, DEC +54:11:36.3 using 35~s exposures with a 2$\times$2 pixel binning. To get colour information, the filter wheel was set to rotate between each exposure, not including the $U$-band. A total of 422 images were taken: 105 each for the $B$ and $V$ filters, and 106 each for the $R$ and $I$ filters.

To reduce these data, we first needed to subclass \texttt{Instrument} to define the MEXMAN-\phoptic{} interface (Section \ref{sec: instruments}). The MEXMAN instrument subclass is given in Listing \ref{listing: MEXMAN instrument}.

\begin{lstlisting}[caption={\texttt{Instrument} subclass for the MEXMAN instrument.},language=Python,label={listing: MEXMAN instrument}]
from astropy import units as u
from astropy.coordinates import EarthLocation
from astropy.time import Time

from phoptic import Instrument


class MEXMAN(Instrument):
    
    def __init__(
        self,
        ):
        
        return super().__init__(
            diameter=0.84 * u.m,
            location=EarthLocation.from_geodetic(
                lon=-115.463611 * u.deg,
                lat=31.044167 * u.deg,
                height=2790 * u.m,
                ),
            pixel_scales={
                'MEXMAN': 0.24645,
                },
            dateobs_kw='JD',
            binning_kw='CCDSUM',
            )


    def get_camera(
        self,
        file = None,
        header = None,
        ):
        
        return 'MEXMAN'


    def get_mjd(
        self,
        file = None,
        header = None,
        ):
        
        if file is not None:
            header = file.get_header()
        
        jd = header[self.dateobs_kw]
        mjd = Time(jd, format='jd').mjd
        
        return mjd


    def get_relative_scintillation_noise(
        self,
        file = None,
        header = None,
        C = 1.67,
        H = 8000 * u.m,
        ):
        
        return super().get_relative_scintillation_noise(file=file, header=header, C=C, H=H)
\end{lstlisting}

We then used this \texttt{MEXMAN} \texttt{Instrument} to reduce this observation. We performed bias and flat-field corrections using the respective routines (Section \ref{sec: corrections}), but omitted dark noise corrections. While observing, MEXMAN's detector is cooled to $\sim 163$~K, and so we assume its dark current is negligible. To estimate image backgrounds, we used \phoptic{}'s default background estimation routine (Section \ref{sec: background}) with a \texttt{box\_size} parameter value of 32; to identify sources, we used \phoptic{}'s default source finder routine (Section \ref{sec: source identification}) with an \texttt{n\_pixels} parameter value of 64. All other reduction parameters were left to their default values.

We constructed source catalogues using the default parameter values (Section \ref{sec: source catalogues}), and computed light curves using aperture photometry with the default aperture (Section \ref{sec: photometry}). In Listing \ref{listing: MEXMAN reduction}, we provide the full reduction script for this observation.

\begin{lstlisting}[caption={Data reduction script for the MEXMAN observation described in Section \ref{sec: MEXMAN example}.},language=Python,label={listing: MEXMAN reduction}]
import os
from pathlib import Path

os.environ['OMP_NUM_THREADS'] = '1'

import phoptic

from instrument import MEXMAN


DATA_DIR = Path('/mnt/SATA1/Data/MEXMAN')
OUT_DIR = Path('Reduced')


def reduce():
    instrument = MEXMAN()
    
    bias_corr = phoptic.BiasCorrector(
        out_directory=OUT_DIR / 'Calibrators',
        data_directory=DATA_DIR / 'biases',
        instrument=instrument,
        )
    
    flat_corr = phoptic.FlatFieldCorrector(
        out_directory=OUT_DIR / 'Calibrators',
        data_directory=DATA_DIR / 'flats',
        instrument=instrument,
        bias_corrector=bias_corr,
        )
    
    background = phoptic.DefaultBackground(32)
    finder = phoptic.DefaultFinder(n_pixels=64)
    
    reducer = phoptic.Reducer(
    out_directory=OUT_DIR,
    data_directory=DATA_DIR / 'object',
    instrument=instrument,
    bias_corrector=bias_corr,
    flat_corrector=flat_corr,
    remove_cosmic_rays=True,
    show_plots=False,
    background=background,
    finder=finder,
    )
    
    reducer.create_catalogs()
    
    photometer = phoptic.AperturePhotometer()
    reducer.photometry(photometer)


if __name__ == '__main__':
    reduce()
\end{lstlisting}

\begin{figure}
    \centering
    \includegraphics[width=\columnwidth]{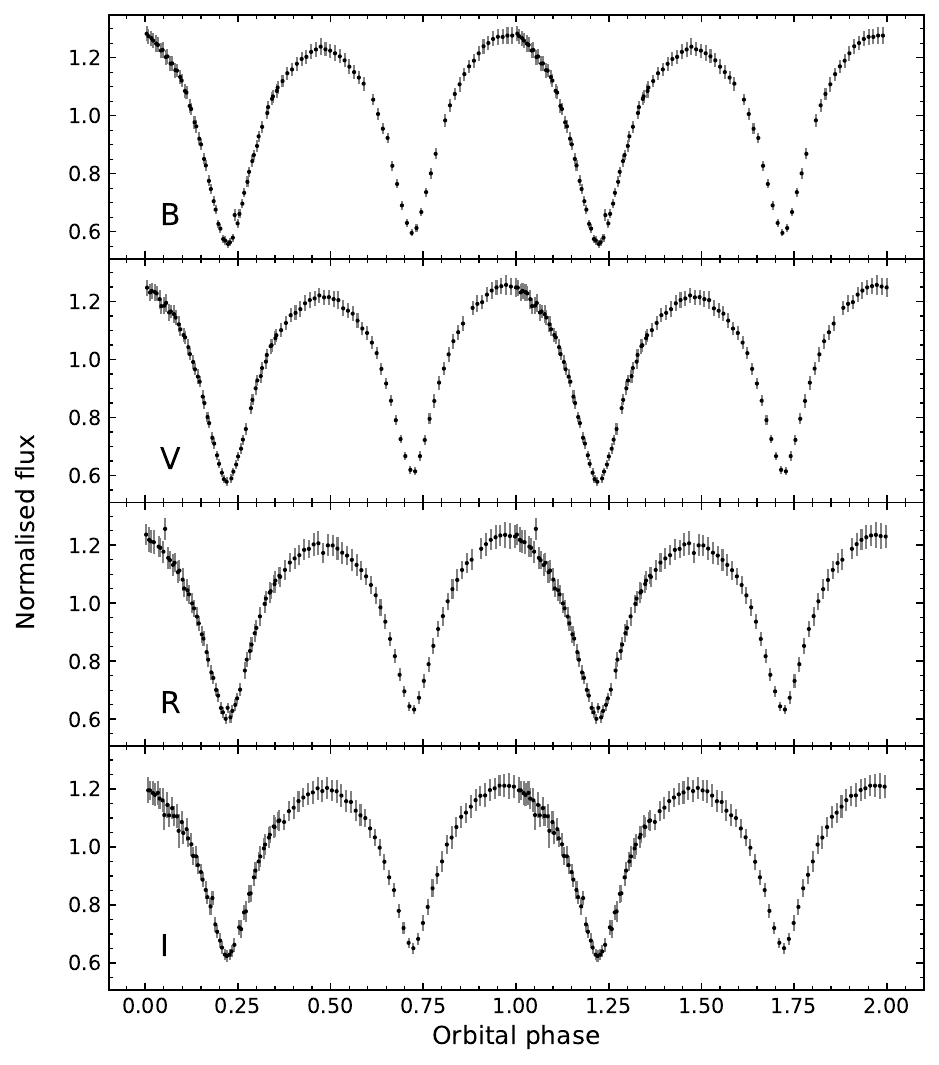}
    \caption{MEXMAN light curves of [GGM2006] 2532136 \citep{Gettel2006} folded on the 6.94~hr orbital period. Two periods are shown for clarity.}
    \label{fig: MEXMAN light curves}
\end{figure}

After reducing these data, we computed relative light curves using an unlisted, non-variable source at RA 10:04:28.1, DEC +54:11:13.0. In Figure \ref{fig: MEXMAN light curves}, we present the resulting light curves folded on [GGM2006] 2532136's 6.94~hr orbital period \citep[][]{Gettel2006}. The time of zero phase was arbitrarily set to the time of the first exposure: BMJD 61074.1770.

\section{Performance}\label{sec: performance}

In this section, we review the performance of \phoptic{} using the following system:
\begin{itemize}
    \item Processor: 2 $\times$ Intel Xeon Gold 6426Y.
    \item Memory: $8 \times 16$~GB 4800~MHz DDR5.
    \item Operating System: Ubuntu 22.04.5 LTS.
\end{itemize}

To speed up data reduction, \phoptic{} leverages the \texttt{multiprocessing} package, which can result in close-to-linear performance scaling on multi-core systems \citep[e.g.,][]{Singh2013}. Specifically, \texttt{multiprocessing} is used to parallelise image alignment and photometry in the \texttt{create\_catalogs()} and \texttt{photometry()} methods of \texttt{Reducer}, respectively. The number of parallel processes created by \texttt{multiprocessing} can be set by passing the desired value to the \texttt{number\_of\_processors} parameter of \texttt{Reducer}. By default, \texttt{number\_of\_processors} will be set to half the number of available logical processors.

\begin{figure*}
    \centering
    \includegraphics[width=\textwidth]{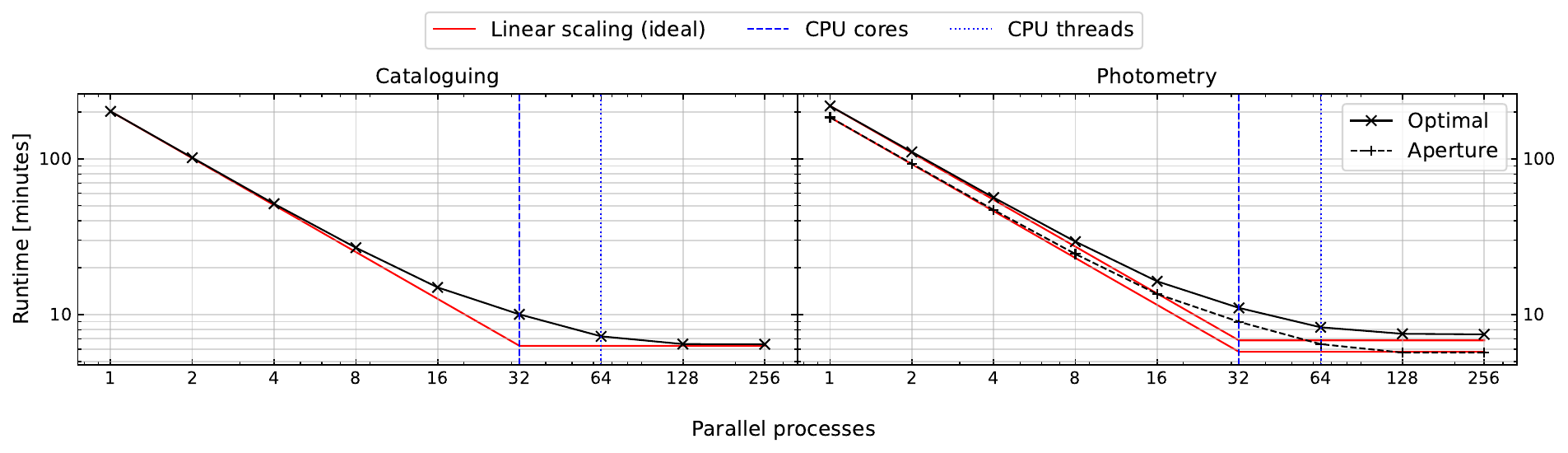}
    \caption{The runtime (in minutes) of the \texttt{Reducer.create\_catalogs()} (left) and \texttt{Reducer.photometry()} (right) methods as a function of the number of parallel processes. The solid red line represents linear (i.e., ideal) scaling up to the number of CPU cores, while the dashed and dotted blue lines show the number of CPU cores and CPU threads in our benchmark system, respectively.}
    \label{fig: benchmark}
\end{figure*}

In Figure \ref{fig: benchmark}, we present the runtime of the \texttt{Reducer.create\_catalogs()} and \texttt{Reducer.photometry()} methods as functions of the number of parallel processes. To perform photometry, we used both the \texttt{AperturePhotometer} and \texttt{OptimalPhotometer} classes with their default parameters (Section \ref{sec: photometry}). For this test, we used 3501 OPTICAM images per camera taken in the 2$\times$2 binning mode. In addition to passing the specified number of processes to the \texttt{Reducer} instance, we also passed \texttt{remove\_cosmic\_rays=True}. All other \texttt{Reducer} parameters were left to their default values.

As can be seen from Figure \ref{fig: benchmark}, the \texttt{Reducer.create\_catalogs()} and \texttt{Reducer.photometry()} methods scale almost linearly when the number of parallel processes is less than the number of CPU cores. As the number of parallel processes approaches the number of cores, however, the scaling diminishes. Nonetheless, it can be seen from this figure that performance continues to scale even after the number of parallel processes exceeds the number of CPU cores. Compared to running a single process, the \texttt{create\_catalogs()} method was 31.2 times faster when running two parallel processes per CPU thread; this is very close to the nominal scaling factor of 32. For the \texttt{photometry()} method, aperture photometry saw a factor 32.4 speed-up when running two processes per thread, suggesting that aperture photometry may be slightly I/O limited rather than compute limited, while optimal photometry saw a 29.0 speed-up factor. In addition to being more scalable, it can also be seen that aperture photometry is considerably faster than optimal photometry, as noted by \cite{Naylor1998}. Quantitatively, aperture photometry was 18 per cent faster when both we limited to a single process, but 32 per cent faster when both were running two parallel processes per CPU thread. We therefore recommend aperture photometry for ``quick-look'' reduction. Finally, we note that there is a negligible benefit to running more than two parallel processes per CPU thread.

\subsection{Aperture vs. optimal photometry}\label{sec: aperture vs. optimal photometry}

Above, we found that aperture photometry is considerably faster than optimal photometry. It is therefore instructive to assess whether optimal photometry is worth the additional expense. To this end, we compare the two methods using the RMS values of the relative light curves between pairs of non-variable sources; the photometry method which performs best is thus the one which produces the lowest RMS. We note, however, this method is sensitive to \textit{correlated} (i.e., red) noise, and that the light curves of non-variable sources may exhibit correlated noise due to systematics. For example, we can see from Figure \ref{fig: source catalogue} that scintillation noise contributes significantly to the brightest sources; since scintillation noise results from atmospheric turbulence (Section \ref{sec: error propagation}), its contribution will necessarily be correlated in time. We therefore omit any sources for which scintillation noise contributes to $> 10$ per cent of the total noise.

For this comparison, we focus on the OPTICAM data reduced in Section \ref{sec: V709 Cas example}. We recall that the source catalogues for this observation are shown in Figure \ref{fig: source catalogue}. As can be seen from Figure \ref{fig: noise characterisation}, scintillation noise contributes $> 10$ per cent to the total noise of most of the $g$-, and all of the $r$- and $i$-band sources identified in Figure \ref{fig: source catalogue}. As such, we limited this comparison to $g$-band Sources 10, 12, 13, 14, and 15; we do not include Source 11 due to its proximity to Source 8.

\begin{figure}
    \centering
    \includegraphics[width=\columnwidth]{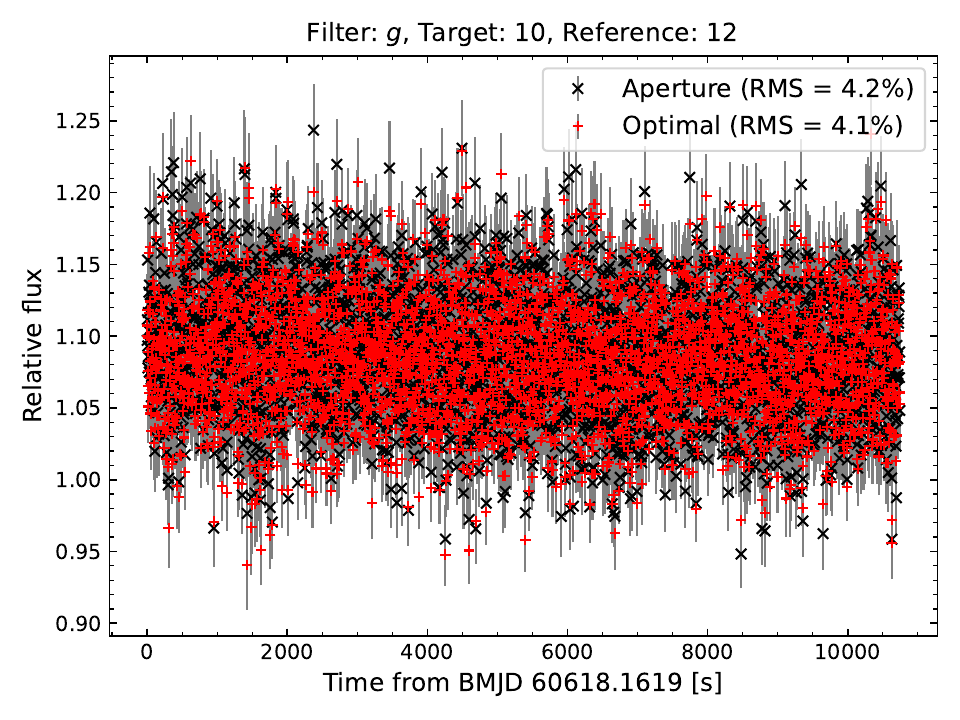}
    \caption{Relative light curves between $g$-band Sources 10 and 12 from the catalogue shown in Figure \ref{fig: source catalogue} as computed using aperture and optimal photometry.}
    \label{fig: aper vs opt example}
\end{figure}

From the above, we found that optimal photometry resulted in an average RMS reduction of just 1.4 per cent. In the worst case, the difference between aperture and optimal photometry was less than 0.1 per cent; in the best case, shown in Figure \ref{fig: aper vs opt example}, optimal photometry reduced the RMS by 3.2 per cent. These differences are considerably smaller than the expected $\sim 10$ per cent \citep{Naylor1998}. We recall, however, that the $\sim 10$ per cent improvement assumes the sky-limited case; as can be seen in Figure \ref{fig: SNR}, all of the sources considered in this comparison have S/N~$\gg 1$, and are therefore not in the sky-limited regime. In this case, optimal photometry therefore provides only a modest improvement in S/N, and may not be worth the additional computational expense in general.

\subsubsection{Noise}\label{sec: noise}

To compare the white and red noise contributions between aperture and optimal photometry, we can use the power spectrum. In the power spectrum, uncorrelated/white noise manifests as high-frequency jitter about a constant power, while correlated/red noise manifests as a low-frequency $f^{- \alpha}$ power law \citep[e.g.,][]{vanderKlis1989}. The average power at high frequencies therefore quantifies the amount of white noise in the underlying light curve, while the low-frequency power law index and normalisation quantifies the amount of red noise.

\begin{figure}
    \centering
    \includegraphics[width=\columnwidth]{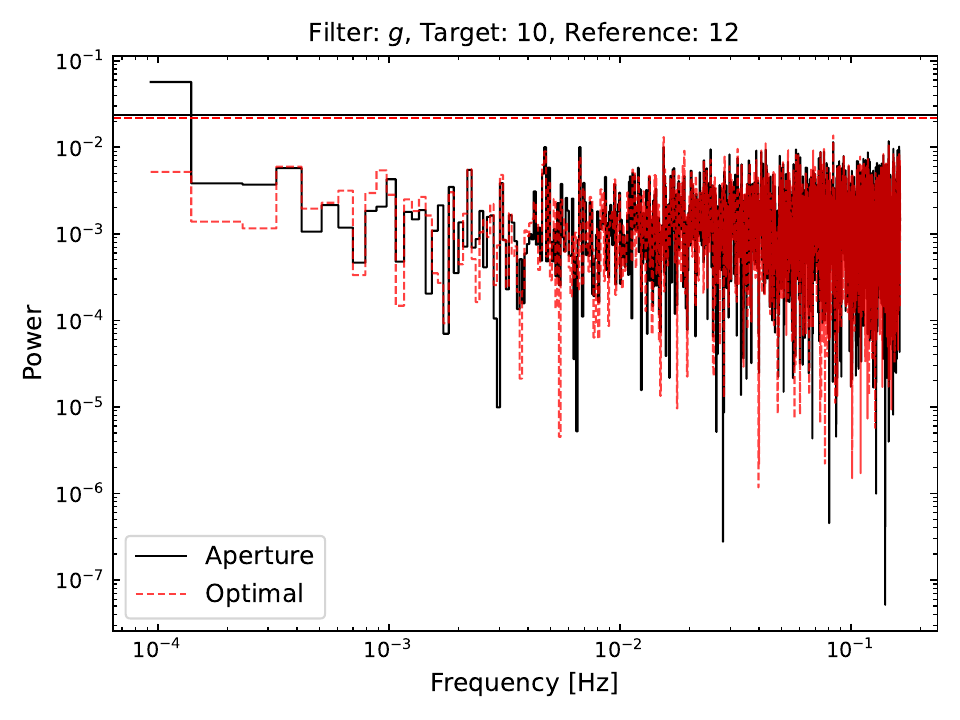}
    \caption{Power spectra of the relative light curves between $g$-band Sources 10 and 12 from Figure \ref{fig: source catalogue} as computed using aperture (solid line) and optimal (dashed line) photometry. The horizontal lines represent the 1 per cent false alarm levels (assuming pure white noise).}
    \label{fig: aper and opt LSP}
\end{figure}

In Figure \ref{fig: aper and opt LSP}, we present the power spectra of the relative light curves between $g$-band Sources 10 and 12 from Figure \ref{fig: source catalogue} as computed using aperture and optimal photometry. We also present 1 per cent false alarm levels for each power spectrum. We computed the false alarm level by randomly shuffling our light curves $10^6$ times, computing the power spectrum of each shuffled light curve, and recording the maximum power of each power spectrum. The 1 per cent false alarm level is then given by the 99th percentile of the distribution of maximum powers. We note, however, that false alarm levels determined this way are only valid if the original light curve does not contain correlated noise \citep[see, e.g.,][]{VanderPlas2018}.

As can be seen from Figure \ref{fig: aper and opt LSP}, both LSPs have similar 1 per cent false alarm levels and average powers, suggesting similar levels of white noise in the two light curves. Moreover, neither LSP exceeds its 1 per cent false alarm level above $2 \times 10^{-4}$~Hz, and so there is no evidence of systematic or intrinsic periodic variability within this frequency range. However, it can be seen from Figure \ref{fig: aper and opt LSP} that the aperture photometry light curve shows evidence of correlated noise: below $\sim 2 \times 10^{-4}$~Hz, the aperture photometry LSP power increases with decreasing frequency - though there are too few points for this to be reliably characterised. In contrast, the optimal photometry LSP appears to be flat across the frequency range. In this case, the amount of correlated systematic noise therefore appears minimal. That said, a full characterisation of OPTICAM's systematics over a range of exposure times is required to understand their noise contributions - both correlated and uncorrelated - which is beyond the scope of this work.

\subsection{Automatic parallelisation}\label{sec: performance caveats}

We note that performance issues may arise due to conflicts between \texttt{multiprocessing} and automatic parallelisation used by libraries such as \texttt{NumPy} \citep{Harris2020}. For the best performance, users should therefore disable automatic parallelisation by inserting the lines:
\begin{lstlisting}[language=Python]
import os
os.environ['OMP_NUM_THREADS'] = '1'
\end{lstlisting}
or the equivalent for their build \textit{before} importing any other libraries. This limits the processes spawned by \texttt{multiprocessing} to a single thread, ensuring the scalable results seen in Section \ref{sec: performance}.

\section{Summary and Future Plans}

\phoptic{} is a simple, modular, and easily extendable Python package that we have developed for performing optical photometry. Initially \phoptic{} was developed as a dedicated pipeline for the OPTICAM instrument; thanks to its flexible instrument interface, however, \phoptic{} can be easily extended to reduce standard-format FITS images from any imaging instrument. The minimal workflow for reducing OPTICAM data with \phoptic{} is as follows:
\begin{enumerate}
    \item First, create a \texttt{DarkNoiseCorrector} instance. Since OPTICAM images report the nominal dark current in their headers, no dark images are required; if \texttt{DarkNoiseCorrector} is instantiated without any dark images, \phoptic{} will attempt to infer the dark noise using the exposure integrated dark current.
    \item Second, create a \texttt{FlatFieldCorrector} instance. The user will need to input the directory path to the raw flat-field images, and the path to the directory in which the resulting master flat-field images will be saved. Additionally, the \texttt{DarkNoiseCorrector} instance created in the previous step may be passed to \texttt{FlatFieldCorrector} to subtract the dark noise from the flat-field images.
    \item Third, create a \texttt{Reducer} instance and pass the \texttt{DarkNoiseCorrector} and \texttt{FlatFieldCorrector} instances to it. The user will need to input the directory path to the raw science images, and the path to the directory in which any output files will be saved. Additionally, the user may optionally pass a number of reduction parameters (Table \ref{tab: reducer params}).
    \item Fourth, call the \texttt{create\_catalogs()} method of \texttt{Reducer} to create source catalogues for each camera (Section \ref{sec: source catalogues}). For sparse fields, it may be necessary to pass \texttt{transform\_type='translation'} and reduce the number of alignment sources.
    \item Finally, create an instance of the desired photometer and pass it to the \texttt{photometry()} method of \texttt{Reducer} to perform photometry (Section \ref{sec: photometry}). The default photometry parameters were chosen with point sources in mind; for extended sources, users will therefore need to define a suitably sized aperture.
\end{enumerate}
To reduce data from other instruments, a custom \texttt{Instrument} instance will be needed (Section \ref{sec: instruments}). Additionally, a bias corrector (Section \ref{sec: bias corrections}) may need to be defined, and the default background and source identification routines (Sections \ref{sec: background} and \ref{sec: source identification}, respectively) may be unsuitable. 

Once data have been reduced, users can compute relative light curves between their source of interest and one or more comparison stars (Section \ref{sec: differential photometry}). After that, users may proceed with their desired analysis, which may include some of the quick-look analyses described in Section \ref{sec: analyzer} or exporting the resulting light curves to \texttt{stingray} for more advanced analyses (Section \ref{sec: stingray}).

\phoptic{} will be a continually evolving project; in the near-future, we plan the following features:
\begin{enumerate}
    \item A ``quick-look'' reduction routine. When observing, it is instructive to be able to quickly compute light curves for targets of interest to ensure that observing parameters, such as exposure time and pixel binning, are suitable. We therefore plan to develop a dedicated quick-look data reduction routine. This routine will remove a lot of \phoptic{}'s flexibility, and minimise the amount of user input required, in order to produce reasonable-quality light curves as quickly as possible. One way of minimising user input will be to implement the algorithm presented in \cite{Broeg2005} to automatically determine suitable comparison sources for performing differential photometry.
    \item A more flexible corrections framework. Currently, \phoptic{} is limited to bias, dark, and flat-field corrections. However, some instruments require bespoke corrections; currently, these that cannot be incorporated directly into \phoptic{}'s reduction. For example, CCD 5 of HiPERCAM exhibits significant fringing (Section \ref{sec: HiPERCAM example}) that can be corrected in a similar way to flat-fielding (Section \ref{sec: flat corrections}). For broader compatibility, we therefore plan to implement a flexible framework for defining and incorporating additional corrections into \phoptic{}.
\end{enumerate}
In addition, we have a number of long-term goals for OPTICAM/\phoptic{}:
\begin{enumerate}
    \item Improved characterisation of OPTICAM's systematics. As noted in Section \ref{sec: OPTICAM intro}, CMOS detectors typically have worse uniformity and stability than CCDs, and feature additional sources of noise. Characterising OPTICAM's systematics will therefore be vital to maximising both the performance of the detectors and the scientific output of the instrument. However, the engineering runs required for this have not yet been performed, and so this is not yet possible.
    \item A generic PSF modelling framework. Currently, \phoptic{} assumes a Gaussian PSF; while this may be a reasonable approximation for many instruments, some instruments may have PSFs that are better approximated using Lorentzian or Moffat profiles. Alternatively, some instruments may have complex PSFs that cannot be adequately described using simple analytical expressions. We therefore plan to implement a flexible PSF modelling framework to allow for the specification of custom PSFs as part of \phoptic{}'s reduction. Custom PSF models may improve optimal photometry performance (Section \ref{sec: photometry}) and allow for more accurate source positioning when manually identifying sources (Section \ref{sec: source picking}).
\end{enumerate}

\section*{Acknowledgements}

We thank the anonymous referees for their constructive comments which have greatly improved the quality of this manuscript.

ZAI gratefully acknowledges support from the UK Research and Innovation's Science and Technology Facilities Council (STFC) grant ST/X508767/1. NCS acknowledges support from STFC grant ST/X001121/1. AC acknowledges support from the Royal Society Newton International Fellowships NF170803, AL\textbackslash221034, and AL\textbackslash24100056. JVHS acknowledges support from STFC grant ST/R000824/1.

This paper is based on observations carried out at the Observatorio Astronomico Nacional on the Sierra San Pedro Martir (OAN-SPM), Baja California, Mexico. We thank Enrique Colorado, Iv\'an Zavala, Joel Herrera, Salvador Zazueta, Francisco Murillo and F\'elix Diaz for their technical support in the continued maintenance of the OPTICAM instrument during its first years of operation at OAN-SPM. We also extend our gratitude to the 2.1 m telescope operators Felipe Montalvo, Francisco Guill\'en, Hortensia Riesgo, and David Rojas for their valuable assistance.

\textbf{The version of \phoptic{} described in this work corresponds to version 0.7.0 \citep{phoptic}.} If \phoptic{} contributes to a publication, please also cite the following projects/works: \texttt{astroalign} \citep{astroalign}; \texttt{astropy} \citep{astropy}; L.A.Cosmic algorithm/\texttt{astroscrappy} \citep{vanDokkum2001, McCully2018}; \texttt{matplotlib} \citep{Hunter2007}; \texttt{numpy} \citep{Harris2020}; \texttt{photutils} \citep{photutils}; \texttt{scipy} \citep{SciPy2020}; \texttt{skimage} \citep{scikit-image}.

\section*{Data Availability}

The raw data used in this work will be made available upon request.

\section*{Conflict of Interest}

\textbf{The authors declare no conflict of interest.}



\bibliographystyle{rasti}
\bibliography{example} 






\bsp	
\label{lastpage}
\end{document}